\documentclass[preprint]{aastex}
\usepackage{epsfig}
\usepackage{amstex,amssymb}
\newcommand{\e}[2]{\mbox{$#1 \times 10^{#2}$}}	% Scientific notation
\begin{document}

\title{Intrinsic and Cosmological Signatures in Gamma-Ray Burst Time
Profiles: Time Dilation}

\author{Andrew Lee and Elliott D. Bloom}
\affil{Stanford Linear Accelerator Center, Stanford University,
Stanford, California 94309}
\makeatletter
\email{alee@slac.stanford.edu and elliott@slac.stanford.edu}
\makeatother

%\and

\author{Vah\'{e} Petrosian}
\affil{Center for Space Science and Astrophysics, Varian 302c,
Stanford University, Stanford, CA 94305-4060 \altaffilmark{1}}
\makeatletter
\email{vahe@astronomy.stanford.edu}
\makeatother

\altaffiltext{1}{Also Astronomy Program and Department of Physics.}

\begin{abstract}
The time profiles of many gamma-ray bursts consist of distinct pulses,
which offers the possibility of characterizing the temporal structure
of these bursts using a relatively small set of pulse shape
parameters.  We have used a pulse decomposition procedure to analyze
the Time-to-Spill (TTS) data for all bursts observed by BATSE up
through trigger number 2000, in all energy channels for which TTS data
is available.  We obtain amplitude, rise and decay timescales, a pulse
shape parameter, and the fluences of individual pulses in all of the
bursts.  We investigate the correlations between brightness measures
(amplitude and fluence) and timescale measures (pulse width and
separation) which may result from cosmological time dilation of
bursts, or from intrinsic properties of burst sources or from
selection effects.  The effects of selection biases are evaluated
through simulations.  The correlations between these parameters among
pulses within individual bursts give a measure of the intrinsic
effects while the correlations among bursts could result both from
intrinsic and cosmological effects.  We find that timescales tend to
be shorter in bursts with higher peak fluxes, as expected from
cosmological time dilation effects, but also find that there are
non-cosmological effects contributing to this inverse correlation.  We
find that timescales tend to be longer in bursts with higher total
fluences, contrary to what is expected from cosmological effects.  We
also find that peak fluxes and total fluences of bursts are
uncorrelated, indicating that they cannot both be good distance
indicators for bursts.
\end{abstract}


\section{Introduction}

Many of the signatures of the cosmological time dilation and the
radiation mechanisms of gamma-ray bursts (GRBs) are hidden in the
temporal and spectral characteristics of GRBs.  The subject of this
paper is the analysis of the temporal properties of the bursts, and
the correlations between intensities and timescales.  We use the BATSE
Time-to-Spill (TTS) data, which can give much higher time resolution
than other forms of BATSE data for most bursts.  The advantages and
shortcomings of this data, our decomposition of the time profiles into
pulses, and the evolution of burst characteristics are described in
greater detail in the accompanying paper \cite{lee:2000}.  What
follows is a brief summary.  (See also
\cite{lee:1996,lee:1998,lee:thesis}.)

Many burst time profiles appear to be composed of a series of
discrete, often overlapping, pulses, often with a \emph{fast rise,
exponential decay} (FRED) shape~\citep{norris:1996}.  The different
pulses may represent emission from distinct subevents within the
gamma-ray burst source.  Therefore, it may be useful to decompose
burst time profiles in terms of individual pulses, each of which rises
from background to a maximum and then decays back to background
levels.  We have analyzed gamma-ray burst time profiles by
representing them in terms of a finite number of pulses, each of which
is described by a small number of parameters.

We have used the phenomological pulse model of \cite{norris:1996} to
decompose gamma-ray burst time profiles into distinct pulses.  In this
model, each pulse is described by five parameters with the functional
form
\begin{equation}
I(t)\  =\  A \exp{\left(-\left\vert\frac{t
 - t_{\text{max}}}{\sigma_{r,d}}\right\vert^{\nu}\right)}\ ,
\label{eq:pulse}
\end{equation}
where $t_{\text{max}}$ is the time at which the pulse attains its
maximum, $\sigma_{r}$ and $\sigma_{d}$ are the rise and decay times,
respectively, $A$ is the pulse amplitude, and $\nu$
(the~``peakedness'') gives the sharpness or smoothness of the pulse at
its peak.

We have developed an interactive pulse-fitting program to perform this
pulse decomposition on the BATSE TTS data. and used this program to
fit pulses to all gamma-ray bursts in the BATSE 3B
catalog~\citep{batse:3b} up to trigger number 2000 in all of the four
BATSE LAD energy channels for which TTS data is available and shows
time variation beyond the normal Poisson noise for the background.  We
fit each channel of each burst separately.  We have obtained 574 fits
for 211 bursts, with a total of 2465 pulses.

In this paper, we focus on the possibility of distinguishing between
intrinsic signatures in the temporal characteristics and those which
arise from their cosmological distribution.  A prominent example of
this is the cosmological time dilation effect, which we expect to see
since some, and possibly all, gamma-ray bursts originate at
cosmological distances.

All timescales in GRBs will be lengthened by a factor of $1 + z$ where
$z$ is the redshift of the burst, as a result of cosmological time
dilation~\citep{paczynski:1992,piran:1992}.  However, this seemingly
straightforward test is not simple.  First of all, given the great
diversity in burst time profiles, it is difficult to decide which
timescale is most appropriate for this test.  It seems unlikely that
any particular timescale is approximately the same in all bursts, so
we expect to find time dilation as a statistical effect, rather than
for individual bursts.

Secondly, redshifts are known only for a few bursts, so that for the
vast majority of bursts we need to use another measure of distance or
redshift.  Most past analyses have used some measure of apparent GRB
brightness for this purpose with the tacit assumption that the
corresponding intrinsic brightness is a standard candle or has a very
narrow distribution.

The observed apparent brightnesses of bursts are generally measured
using either peak fluxes, which give the instantaneous intensity of
bursts when they peak, or fluences, which measure the total output of
bursts integrated over their entire durations.  The brightness
measures can also be divided another way, into photon measures and
energy measures.  Thus, there are several different measures of the
apparent brightnesses of bursts.  The BATSE burst catalogs give peak
photon fluxes and total energy fluences for bursts.  The pulse-fitting
data presented here can be used to determine count fluxes and count
fluences.  Most previous work on the evidence for time dilation in
burst time profiles has binned the bursts into two or three brightness
classes using the peak flux as a measure of brightness, and compared a
measure of total burst duration these classes.  Use of fluence as a
brightness measure has been promoted by \cite{petrosian:1996} and
\cite{lloyd:1999}.

In this paper, we use a number of different timescale and brightness
measures.  We will describe their correlations using power laws.
Although cosmological models generally predict more complex
relationships than a simple power law, it would be fruitless to
attempt to fit anything more complex than a power law using the
pulse-fitting data, which appears to have a large intrinsic scatter.
To contrast the cosmological versus the intrinsic signatures, we
compare the relations or correlations between strengths and timescales
among bursts, which should contain the signatures of cosmological time
dilations, with the same correlations among pulses of individual
bursts, which can only contain the intrinsic effects.  It is likely
that some of these correlations are affected by selection effects in
our fitting procedures.  To investigate the importance of these, we
have carried out extensive simulations which are described in the
accompanying paper \cite{lee:2000}.  We use the results of these
simulations to test whether or not the correlations we find are
properties of the bursts or are products of our procedures.  In the
next section, we define the various timescales and burst strengths
used in this analysis.  The correlations relevant to the ``time
dilation'' tests are discussed in Section~\ref{sec:timedilation} and
the correlations between other quantities within bursts and among
bursts are described in Section~\ref{sec:othercorr}.  In
Section~\ref{sec:discuss} we discuss the significance of these
correlations.

It should be noted that many of the simulated bursts were affected by
a truncation that almost never occurred in the actual BATSE TTS data.
The TTS data is truncated at $2^{20}$ counts or 240 seconds, whichever
occurs first.  In nearly all of the actual bursts, the 240 second
limit is reached first, while in many of the the simulated bursts, the
$2^{20}$ count limit is reached first.  This truncation can shorten
the observed time intervals between the first and last pulses in a
burst, and between the two highest amplitude pulses in a burst, but
not the observed pulse widths or the observed time intervals between
consecutive pulses.  \emph{Therefore, all discussions of the first two
kinds of time intervals in simulated bursts will only consider
simulated bursts where no pulses were truncated by the $2^{20}$ count
limit.}

\section{Timescales and Intensities}

We now describe the characteristics used in our correlation studies
and the selection and procedural biases associated with each of them.

\subsection{Intensities}

We use peak count rates and count fluences as measures of burst
intensity or strength.  For individual pulses, the peak count rate is
given by the amplitude $A$ and the count fluence by
\begin{equation}
\mathcal{F} = A \int_{-\infty}^{\infty}{I(t)dt} = A \frac{\sigma_{r} + \sigma_{d}}{\nu}\Gamma\left(\frac{1}{\nu}\right) .
\label{eq:area}
\end{equation}
where $\Gamma$ is the gamma function.  For a burst, on the other hand,
the peak count rate is $A_{\text{max}}$, the largest amplitude of the
pulses in the burst, and the total count fluence is $\mathcal{F} =
\sum{\mathcal{F}_{i}}$, summed over all pulses.

\subsection{Time Intervals Between Pulses}

The most obvious timescale for individual pulses is the \emph{pulse width},
which is given by
\begin{equation}
T_{f} = A (\sigma_{r} + \sigma_{d}) (-\ln f)^{\frac{1}{\nu}} .
\label{eq:fwfm}
\end{equation}
where $f$ is the fraction of the peak height at which the width is
measured. and $\nu$ is the ``peakedness'' parameter.  In this paper,
we use the case $f = 1/2$, for which the width is the full width at
half maximum (FWHM).  We will discuss the correlations between pulse
width and intensity measures in the next section.  Here we consider
some other timescales, namely the \emph{time intervals between
pulses}, which may also be characteristic of the gamma-ray production
mechanisms.  There are several possible choices of time intervals.
We'll examine the \emph{intervals between consecutive pulses} first,
which may have the following selection effect: Two pulses with short
separations between their peaks may have a large overlap, and thus be
identified as only one pulse.  This will limit the shortest interval
between pulses, introducing a selection bias.  On the other hand, when
two pulses have a long separation between them, additional smaller
pulses may be resolved between them that wouldn't be resolved if the
separation were smaller.  This will limit the the longest intervals
between consecutive pulses, introducing another selection bias.

\begin{figure}\plotone{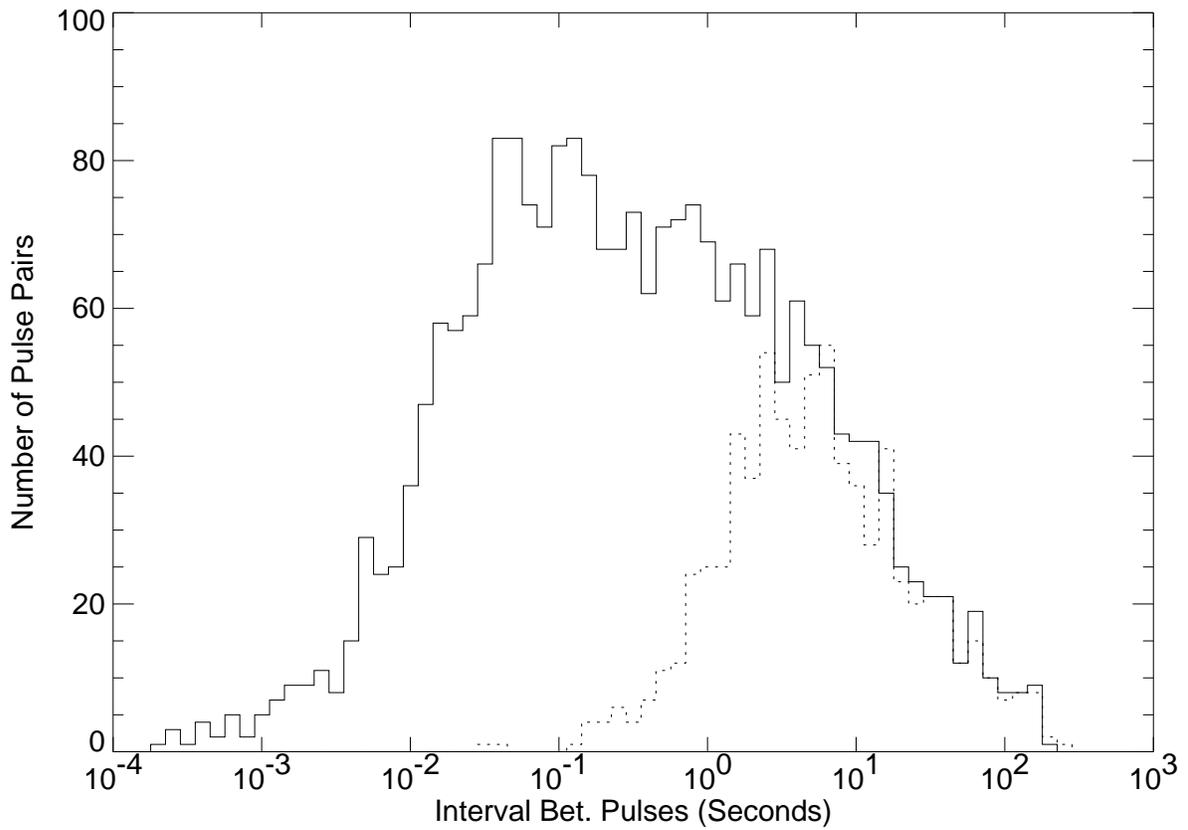}\caption{Distribution of intervals between peak times of adjacent pulses in initial simulations (solid histogram) and in the results of the fits to the simulated data (dashed histogram).  Note that a large number of bursts with small separations are combined
with nearby stronger pulses.  \label{sim_interval}}
\end{figure}

Figure~\ref{sim_interval} shows the distributions of the intervals
between the peak times $t_{\text{max}}$ of adjacent pulses for the
simulations and the fits to simulations.  It shows that the fitting
procedure identifies pulses with longer separations correctly, but
misses most pulses with shorter separations.

\begin{figure}
\epsfig{file=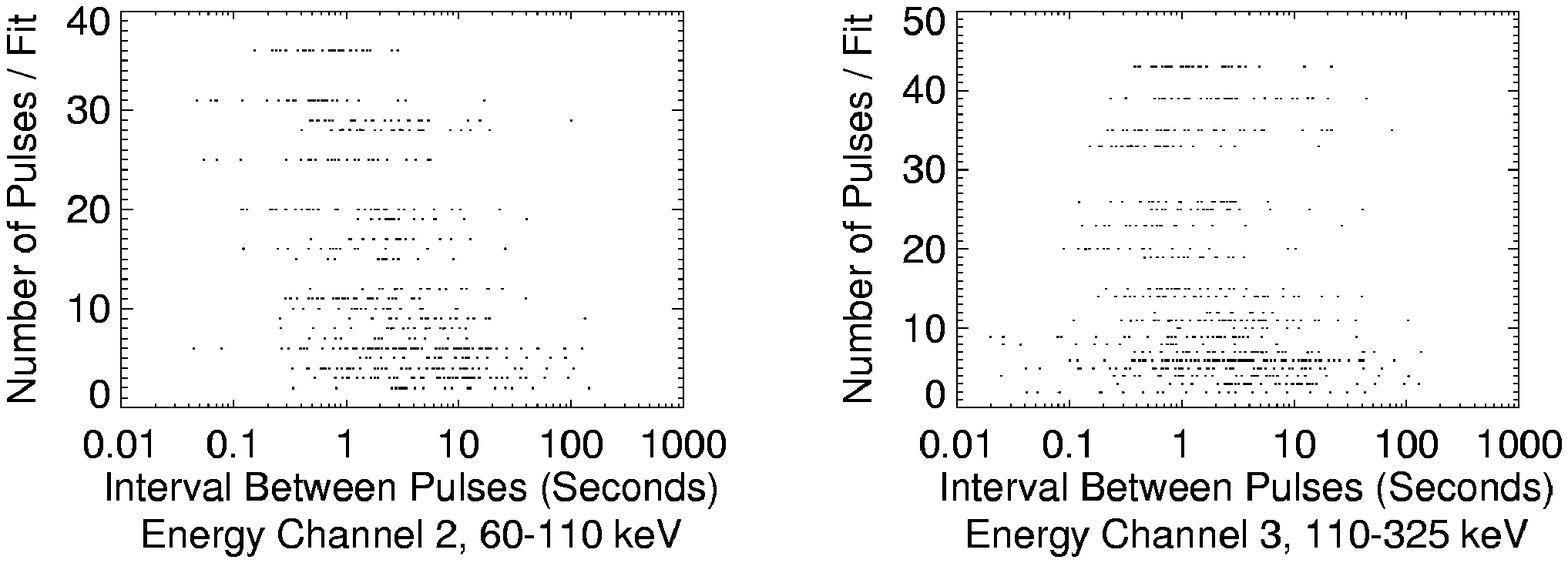}
\epsfig{file=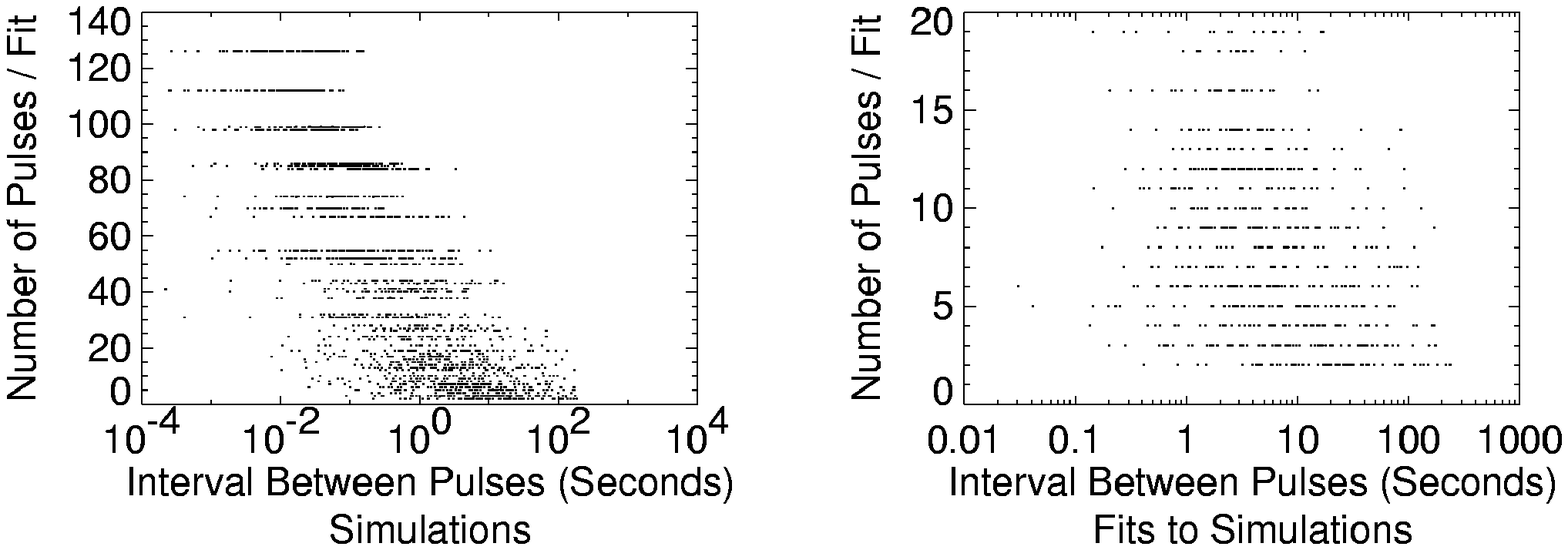}\caption{Number of pulses per burst versus intervals between adjacent pulses for BATSE energy channels~2
and 3 (upper panels) and for initial simulated bursts and fits to
these bursts (lower panels).  Similar results were obtained for
channels~1 and 4, which have much fewer pulses. \label{npvsdt}}
\end{figure}

Figure~\ref{npvsdt} shows the time intervals between consecutive
pulses for bursts with different numbers of pulses, as derived from
our fits to the BATSE data and from the simulations.  Note that here
and in similar figures to follow, we show only data from channels~2
and 3.  In general, channels 1 and 4 show similar behavior, but
results from these channels have lower significance because these
channels contain fewer pulses.  Table~\ref{tab:npvsint}, columns~(a)
gives the Spearman rank-order correlation coefficients $r_{s}$ between
these two quantities, and the probabilities that the observed
correlations have occured by chance.  These show that pulses tend to
be closer together in bursts with more pulses, in both the actual
bursts, and in the simulated bursts and in the fits to simulated
bursts.  One selection effect that may contribute to this result in
actual bursts is that more complex bursts may simply be bursts with
stronger signal-to-noise ratios, which allows more pulses to be
resolved within any given time interval.  Our analysis of the
simulated bursts and the fits to the simulations show similar
results. this result is as expected, since pulse peak times were
generated independently of each other and of the number of pulses per
burst, so more complex bursts will tend to have more pulses in any
given time interval.  The correlation is weaker for the fits to
simulated bursts than for the original simulated bursts because the
fitting procedure tends to miss pulses with shorter separations.

\begin{deluxetable}{crlrlrl}
\tablecaption{Correlation Between Number of Pulses per Burst and Intervals Between (a) Adjacent, (b) First and Last, and (c) Two Highest Amplitude Pulses. \label{tab:npvsint}}
\tablehead{
\colhead{Energy} & \multicolumn{2}{c}{(a) Adjacent} & \multicolumn{2}{c}{(b) First to Last} & \multicolumn{2}{c}{(c) Two Highest} \\
\colhead{Channel} & \colhead{$r_{s}$} & \colhead{Prob.} & \colhead{$r_{s}$} & \colhead{Prob.} & \colhead{$r_{s}$} & \colhead{Prob.}}
\startdata
1 & -0.39 & \e{1.8}{-14} & 0.53 & \e{3.6}{-8} & -0.07 & 0.50 \\
2 & -0.47 & \e{3.0}{-33} & 0.50 & \e{2.7}{-8} & -0.12 & 0.23 \\
3 & -0.24 & \e{1.6}{-10} & 0.56 & \e{4.4}{-11} & 0.14 & 0.13 \\
4 & -0.27 & \e{8.0}{-5} & 0.52 & 0.0013 & -0.16 & 0.35 \\
Sim. & -0.80 & 0 & 0.55 & \e{1.1}{-14} & 0.01 & 0.86 \\
Fits to Sim. & -0.35 & \e{6.5}{-23} & 0.49 & \e{6.2}{-10} & -0.10 & 0.22 \\
\enddata
\end{deluxetable}

Another time interval, \emph{the interval between the peak times of
the first and last pulses} in a burst, might be expected to give a
good measure of the \emph{total duration} of the burst.  However, the
determination of this interval can be greatly affected by whether or
not low amplitude pulses can be identified above background.  This is
essentially the same effect as the sensitivity of the $T_{90}$
interval to the signal-to-noise ratios of
bursts~\citep{norris:1996c,lee:1997}.

\begin{figure}
\epsfig{file=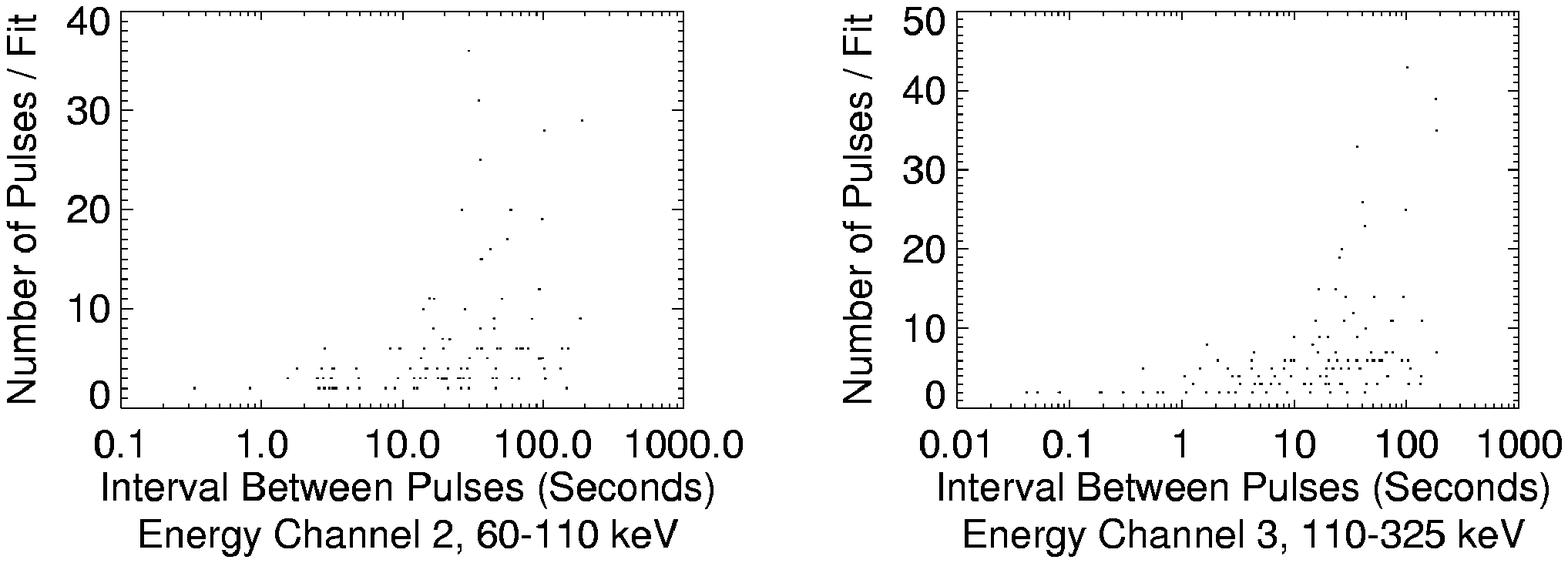}
\epsfig{file=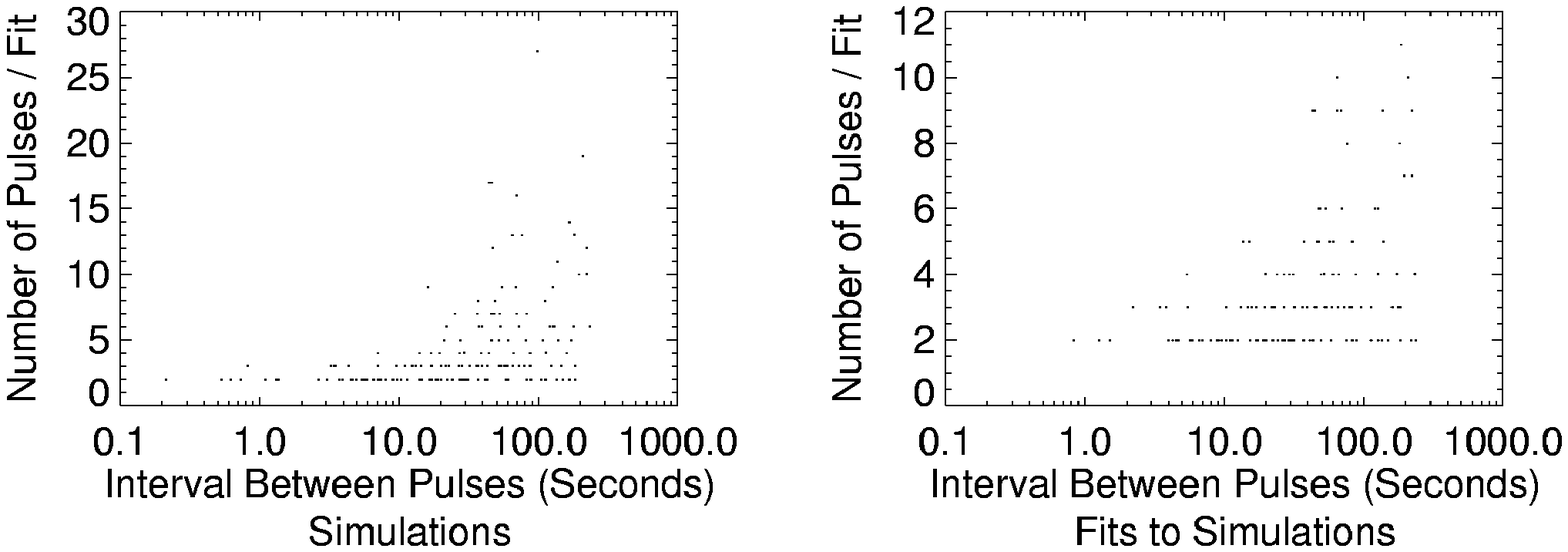}\caption{Same as
Figure~\ref{npvsdt}, except number of pulses per burst versus interval
between first and last pulse in each burst. \label{npvsdur}}
\end{figure}

Figure~\ref{npvsdur} and columns~(b) of Table~\ref{tab:npvsint}
compare the number of pulses in each burst with the time interval
between the first and last pulses in each burst.  They show that the
time intervals between the first and last pulse are greater in bursts
with more pulses, both in actual bursts, and in simulated bursts and
fits to simulated bursts.  In actual bursts, this may result from the
selection effect described above; more complex bursts may simply have
stronger signal-to-noise ratios, making it easier to identify earlier
and later pulses.  In the simulated bursts and the fits to simulated
bursts, this is also as expected since the peak times of pulses were
generated independently of the number of pulses in each burst.

A third time interval, the interval \emph{between the peak times of
the two highest amplitude pulses} in a burst, may also represent a
characteristic time scale for the entire burst.  Determination of this
interval should be less affected by the selection effects that we have
seen with the intervals between consecutive pulses.  However, the
identification of the two highest pulses may be affected by whether a
particular structure in a burst is identified as a single pulse with
large amplitude or as multiple overlapping pulses with smaller
amplitudes.  The interval between the two highest amplitude pulses
should be less influenced by the selection effects in the fitting
procedure that affect the interval between the first and last pulses
in a burst.

\begin{figure}
\epsfig{file=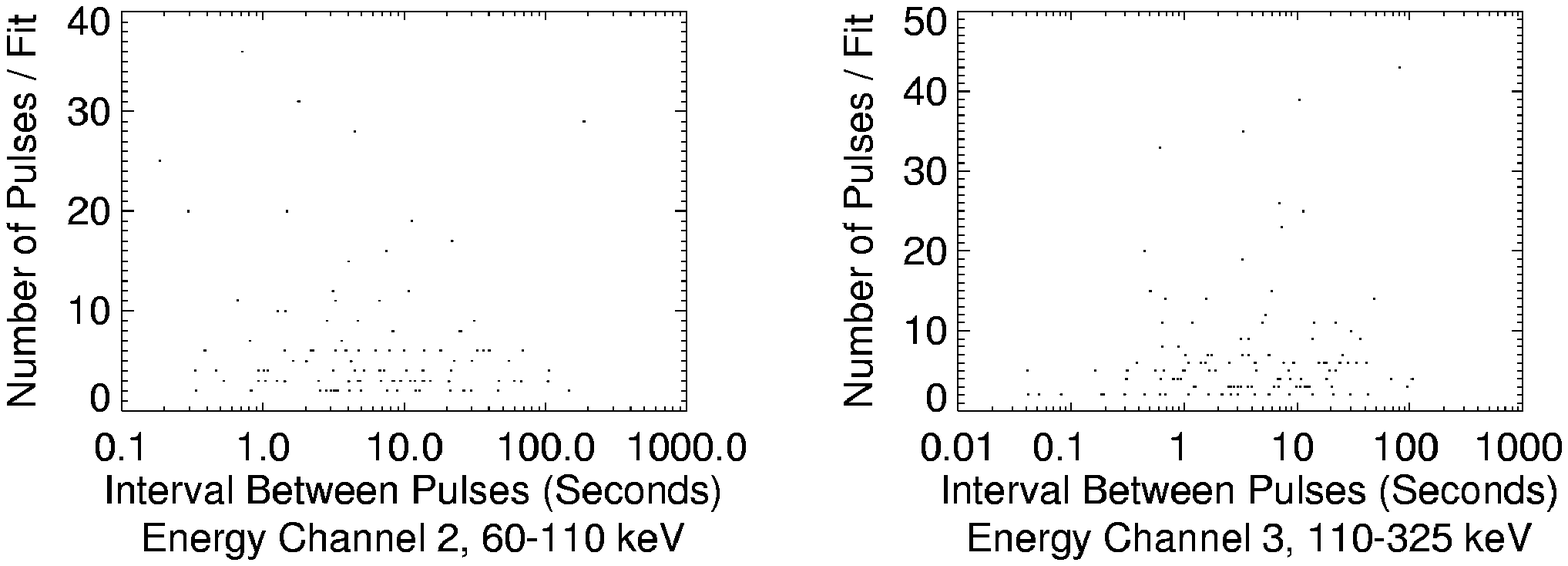}
\epsfig{file=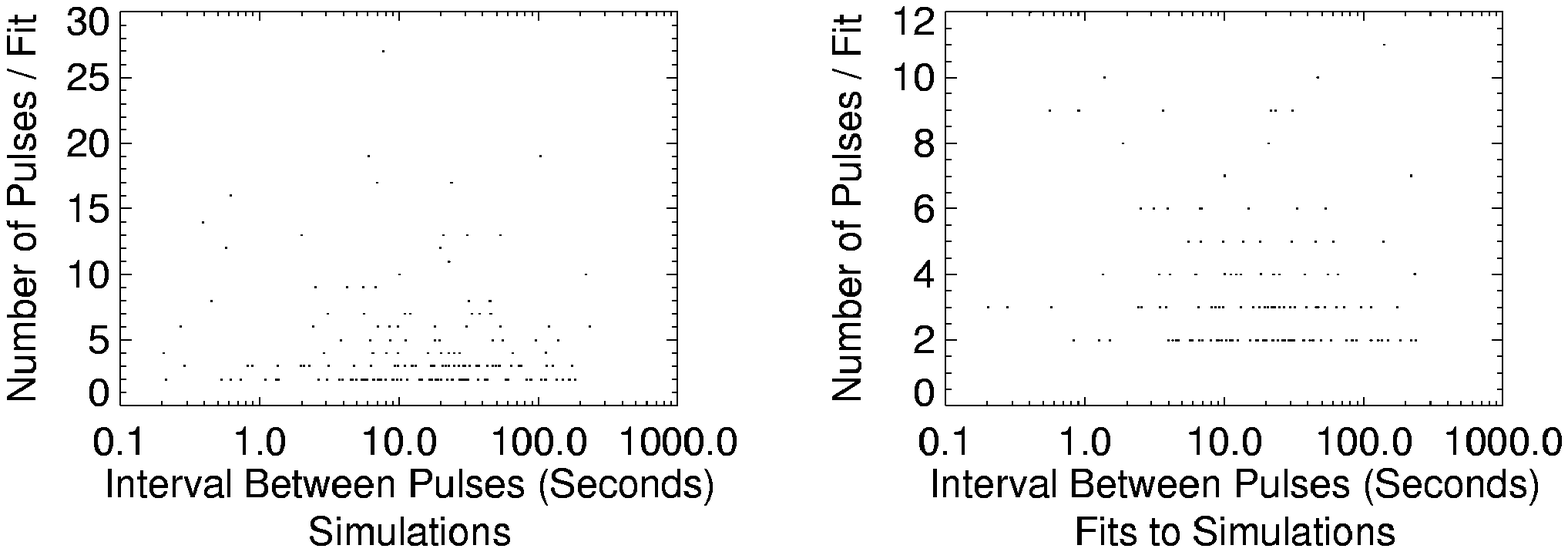}\caption{Same as
Figure~\ref{npvsdt}, except number of pulses per burst versus
intervals between two highest amplitude pulses in each
burst. \label{npvsint}}
\end{figure}

Figure~\ref{npvsint} and Table~\ref{tab:npvsint}, columns~(c) show the
correlations between the number of pulses in each burst and the time
intervals between the two highest amplitude pulses in each burst, both
for actual bursts, and for simulated bursts and fits to simulated
bursts.  It appears that unlike the first two time intervals described
above, there is no tendency for the third time interval to be shorter
or longer in bursts with more pulses.  This suggests that the interval
between the two highest pulses in each fit isn't subject to the
signal-to-noise selection effects that affect both the intervals
between adjacent pulses and the interval between the first and last
pulse in each burst.

The upshot of the above analysis is that the correlations between time
intervals and numbers of pulses per burst (or complexity) in the
simulated bursts is similar to that of the actual BATSE data,
indicating that the simulated data provides a good representation of
these aspects of the actual data, and can be used to determine the
biases in the data and in the fitting procedure.

\section{Time Dilation}
\label{sec:timedilation}

We now consider the correlations between timescales and intensities
among pulses within bursts and among the bursts to determine the
presence of time dilation or time stretching and to test if this is
due to cosmological redshift of the sources.

\subsection{Peak Luminosity as a Standard Candle}

\begin{figure}
\epsfig{file=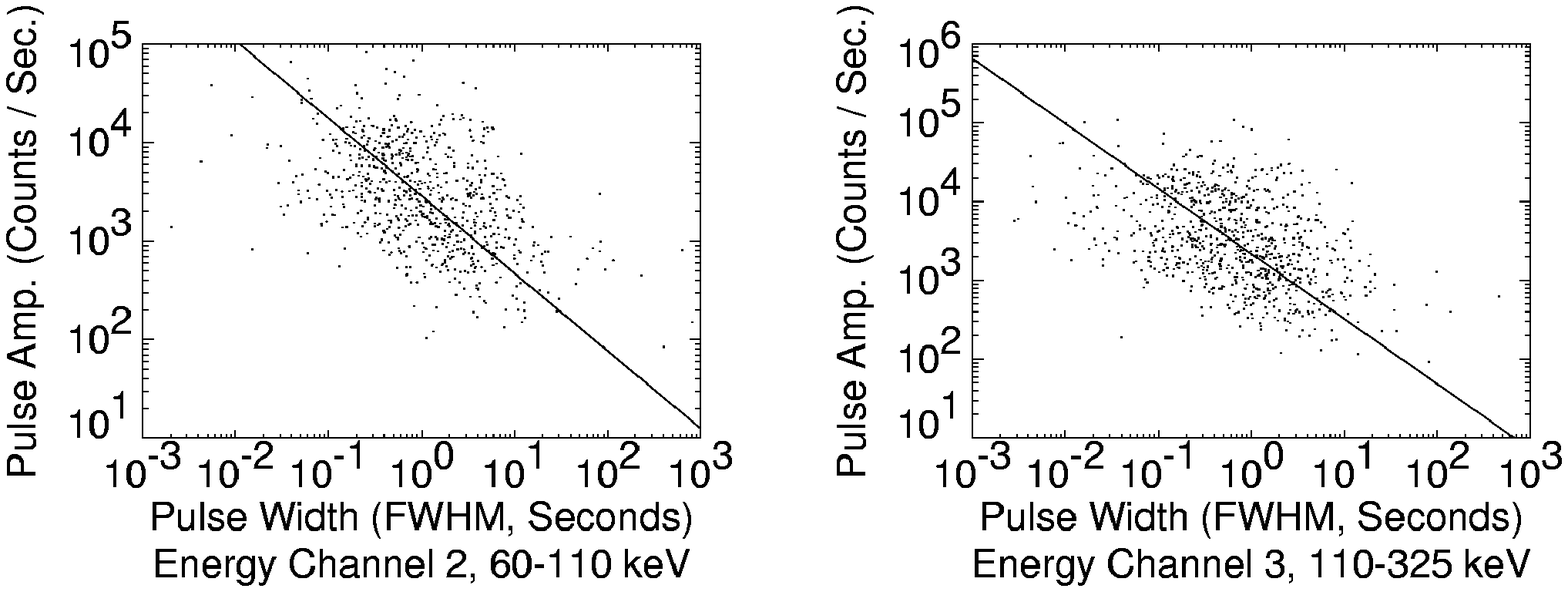}
\epsfig{file=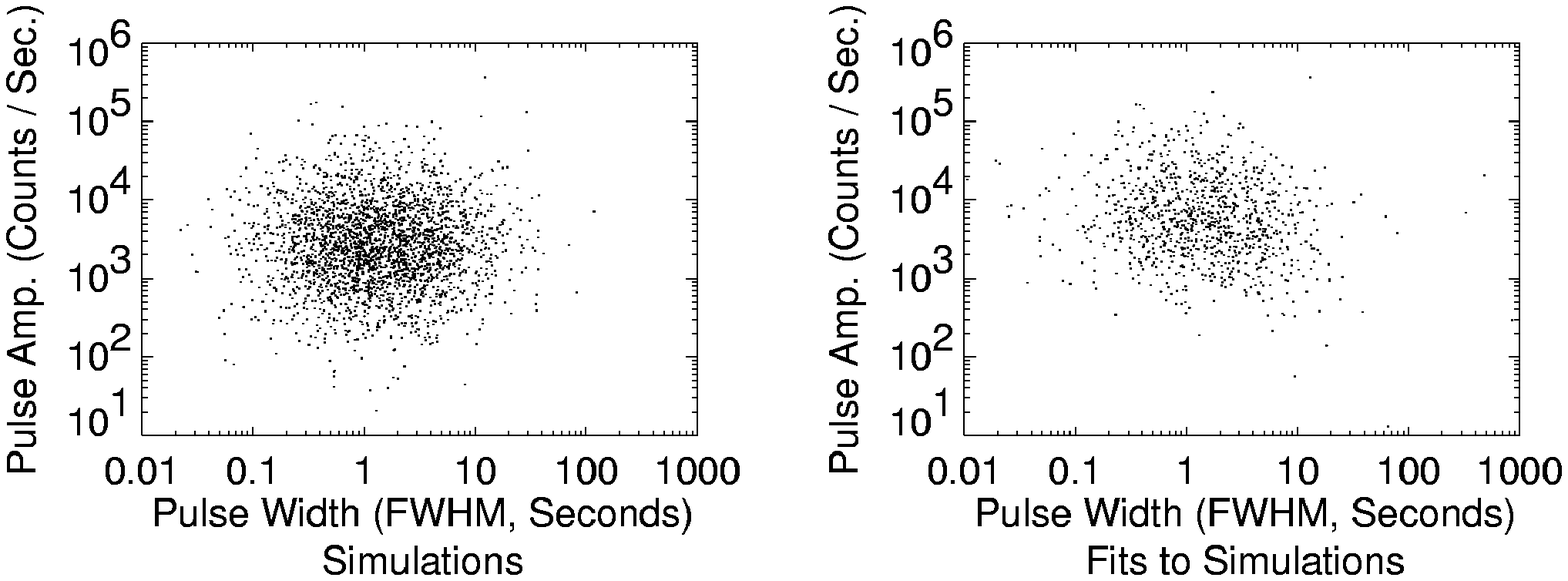}\caption{Pulse amplitude versus pulse width (FWHM) for all pulses in all bursts combined.  The solid lines are obtained from least-squares fits using the OLS bisector method to the logarithms.  In the initial simulations and the fits to the simulations (bottom panels), the correlations were insignificant, so no fits were made.  \label{ampvsw}}
\end{figure}

If we assume that the peak luminosities of bursts are approximately a
standard candle, then the correlations between pulse amplitudes and
timescales can be used to test time dilation.  This corresponds to the
amplitudes of the constituent pulses in bursts.  It has previously
been found that higher amplitude pulses have shorter durations (are
narrower),~\citep{davis:1994,norris:1994b,davis:thesis}, but it has
been noted that this could be in part or entirely an intrinsic
property of bursters.~\citep{norris:1998}.  A potential problem with
using peak flux as a distance measure for bursts observed by BATSE is
that data binned to 64~ms have been typically used, so that the peak
fluxes of bursts with sharp spikes may be underestimated.
(See~\cite{lee:1997}.)  This should be less of a problem with the
variable time resolution TTS data, where the time resolution is
inversely proportional to the count rate and every spill represents
the same number of counts.  The pulse-fitting data from actual BATSE
bursts shown in the upper panels of Figure~\ref{ampvsw} clearly shows
that higher amplitude pulses tend to be narrower, or have shorter
durations.  Table~\ref{tab:ampvsw} gives the Spearman rank-order
correlation coefficients, which show that pulse amplitudes and pulse
widths are inversely correlated in all energy channels.  The table
also gives fitted power laws for pulse amplitude as a function of
pulse width.  These were obtained by applying the ordinary least
squares (OLS) bisector linear regression
algorithm~\citep{isobe:1990,lee:thesis}.

\begin{deluxetable}{crlr}
\tablecaption{Correlation Between Pulse Amplitude and Pulse Width (FWHM) for All Pulses in All Bursts Combined, and the Fitted Power Law Index $\alpha$. \label{tab:ampvsw}}
\tablehead{
\colhead{Energy Channel} & \colhead{$r_{s}$} & \colhead{Prob.} & \colhead{$\alpha$}}
\startdata
1 & -0.53 & \e{9.9}{-39} & $-0.73 \pm 0.03$ \\
2 & -0.49 & 0 & $-0.79 \pm 0.02$ \\
3 & -0.44 & \e{1.3}{-42} & $-0.83 \pm 0.02$ \\
4 & -0.52 & \e{2.4}{-20} & $-0.75 \pm 0.03$ \\
Simulation & 0.0068 & 0.73 & \nodata \\
Fit to Sim. & -0.14 & \e{2.1}{-6} & \nodata \\
\enddata
\end{deluxetable}

The lower panels of Figure~\ref{ampvsw} shows the pulse amplitudes
versus pulse width for all pulses in all simulated bursts combined,
for the initial simulations and for the fits to the simulations.  The
fitting procedure tends to miss lower amplitude pulses, but doesn't
appear to have strong selection effects in pulse width.  However, the
fitting procedure introduces an anticorrelation between pulse
amplitudes and pulse widths, as shown in the last two rows of
Table~\ref{tab:ampvsw}.  By design, there is no correlation between
pulse width and pulse amplitude in the initial simulation, but there
is a negative correlation between pulse width and pulse amplitude in
the fits to the simulations.  However this correlation appear to be
weaker and have far less statistical significance than in the fits to
actual BATSE bursts.

It is difficult to draw concrete conclusions from the correlations in
the combined set of pulses.  To distinguish cosmological from
intrinsic correlations, we should compare the correlations among
bursts and among pulses within individual bursts.

\subsection{Cosmological Effects}

\begin{figure}
\epsfig{file=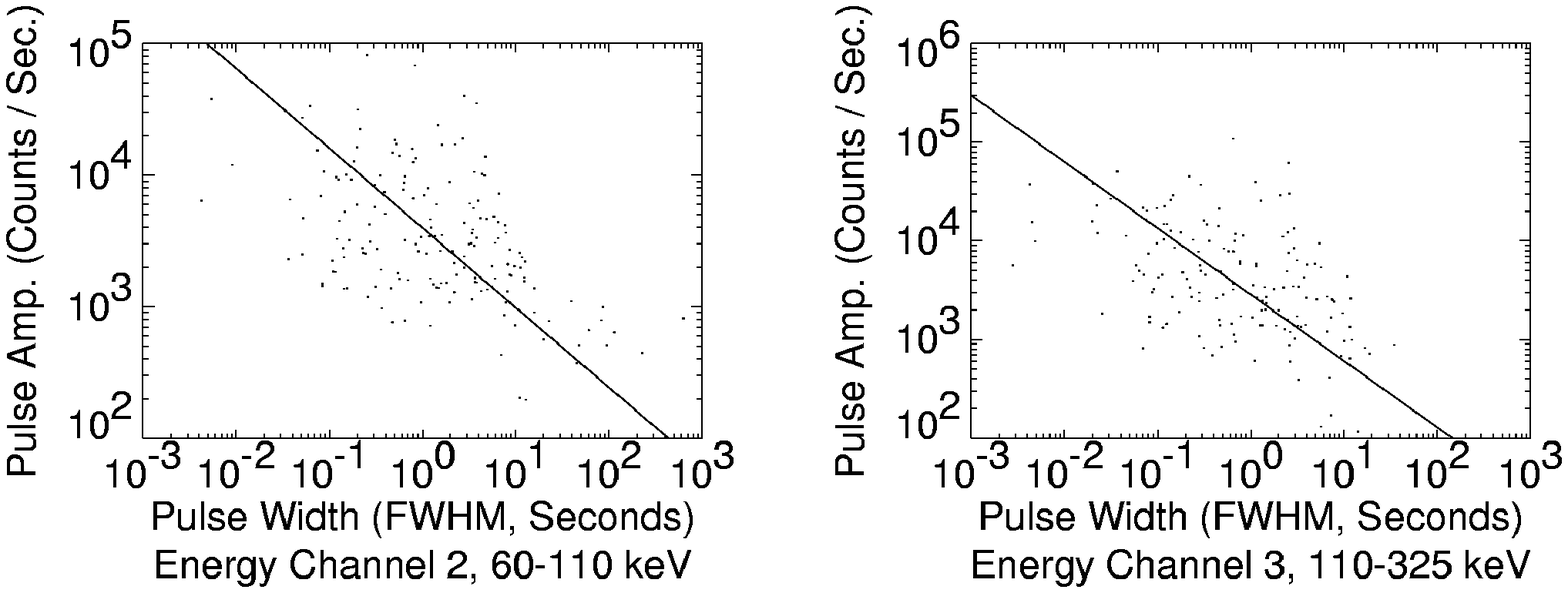}
\epsfig{file=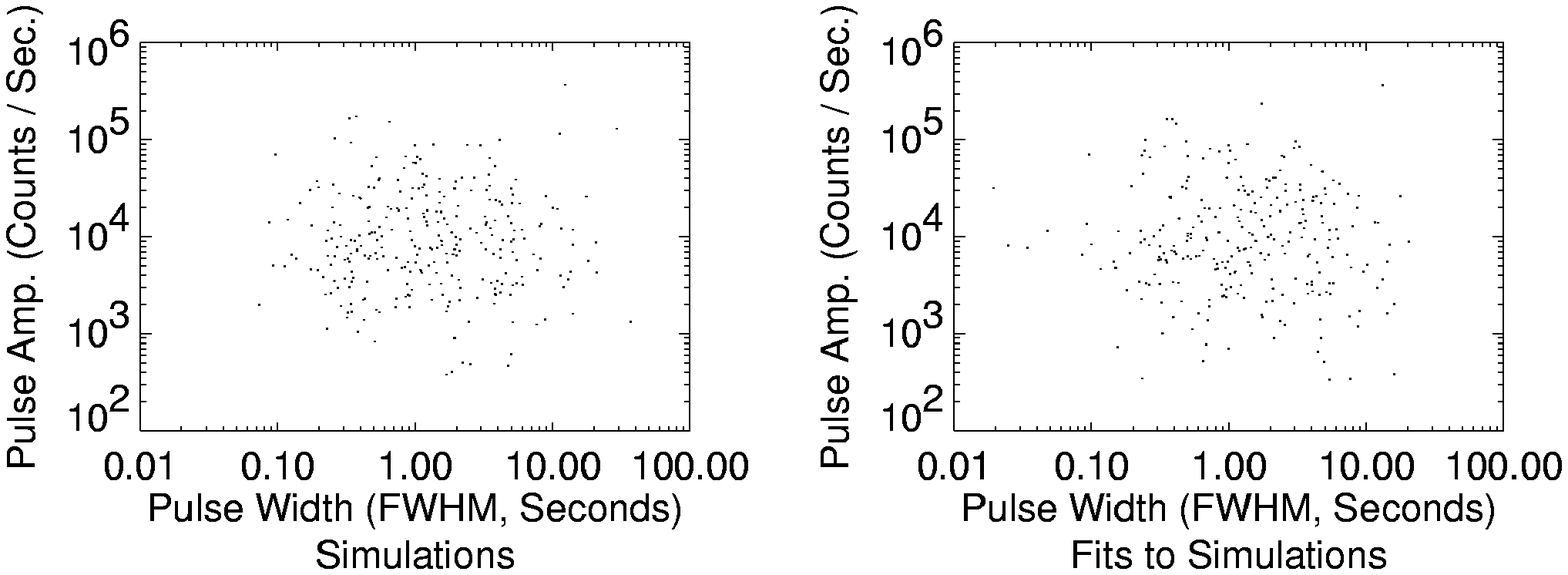}\caption{Same as Figure~\ref{ampvsw}, except for the highest amplitude pulse in each burst.  \label{peak_ampvsw}}
\end{figure}

For testing the first type of correlation, we use the peak fluxes of
each of the bursts, \emph{i.e.}, the amplitudes of the highest
amplitude pulses, and the widths of the same pulses.  These data and
their analysis (shown in Figure~\ref{peak_ampvsw} and columns~(a) of
Table~\ref{tab:peak_ampvswint}) shows a strong inverse correlation
between peak pulse amplitude and pulse width in the actual BATSE
bursts, but not in the simulated bursts or the fits to the simulated
bursts.  This suggests that the correlations observed in the fits to
actual bursts observed by BATSE are not caused by selection effects in
the fitting procedure, so they may arise from cosmological time
dilation, intrinsic properties of the bursters, or selection effects
arising from the BATSE triggering criteria.

\begin{deluxetable}{crlrrlr}
\tablecaption{Correlation Between Highest Pulse Amplitude and (a) Width of Highest Amplitude Pulse and (b) Interval Between Two Highest Pulses in Each Burst, and the Fitted Power Law Index $\alpha$. \label{tab:peak_ampvswint}}
\tablehead{
\colhead{Energy} & \multicolumn{3}{c}{(a) Pulse Width} & \multicolumn{3}{c}{(b) Interval} \\
\colhead{Channel} & \colhead{$r_{s}$} & \colhead{Prob.} & \colhead{$\alpha$} & \colhead{$r_{s}$} & \colhead{Prob.} & \colhead{$\alpha$}}
\startdata
1 & -0.57 & \e{9.1}{-15} & $-0.60 \pm 0.05$ & -0.42 & \e{2.7}{-5} & $-0.86 \pm 0.06$ \\
2 & -0.52 & \e{5.8}{-14} & $-0.61 \pm 0.04$ & -0.42 & \e{7.2}{-6} & $-0.91 \pm 0.06$ \\
3 & -0.51 & \e{1.1}{-12} & $-0.67 \pm 0.05$ & -0.34 & \e{1.8}{-4} & $-0.83 \pm 0.06$ \\
4 & -0.71 & \e{7.1}{-12} & $-0.64 \pm 0.05$ & -0.40 & 0.017 & $-0.82 \pm 0.11$ \\
Sim. & 0.0059 & 0.92 & \nodata & 0.03 & 0.74 & \nodata \\
Fit to Sim. & -0.075 & 0.20 & \nodata & -0.01 & 0.93 & \nodata \\
\enddata
\end{deluxetable}

\subsection{Intrinsic Effects}

\begin{figure}
\epsfig{file=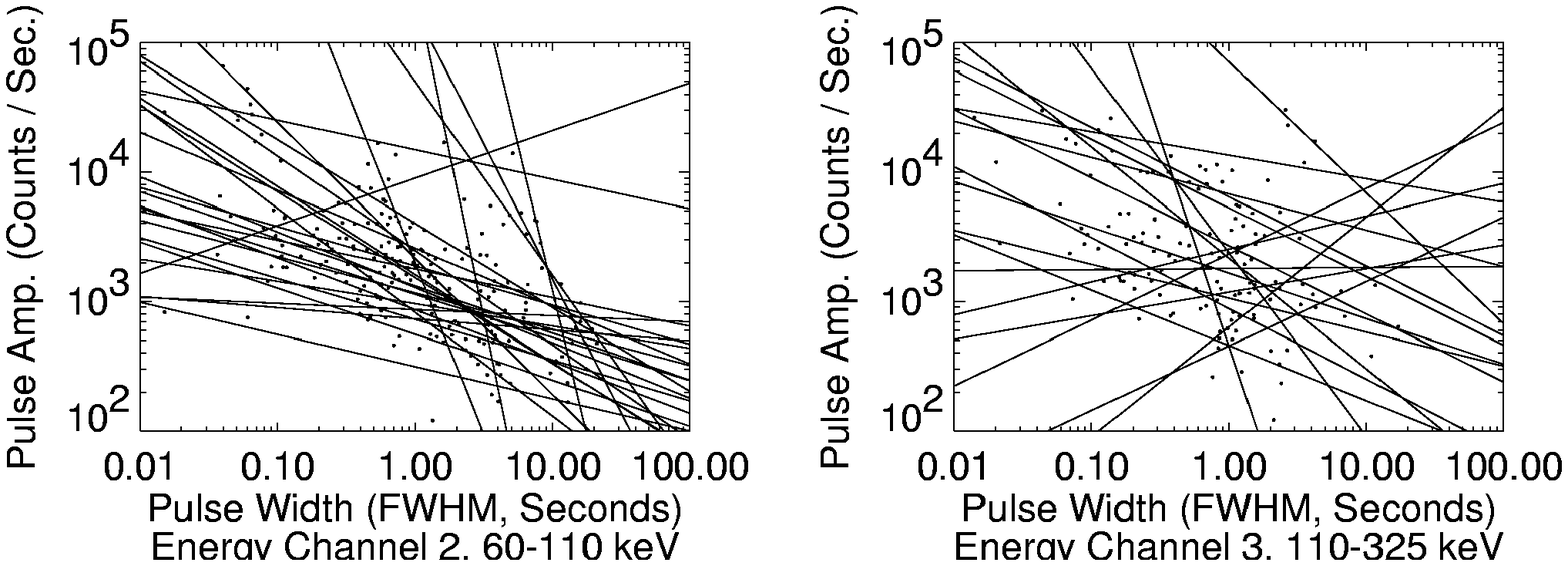}
\epsfig{file=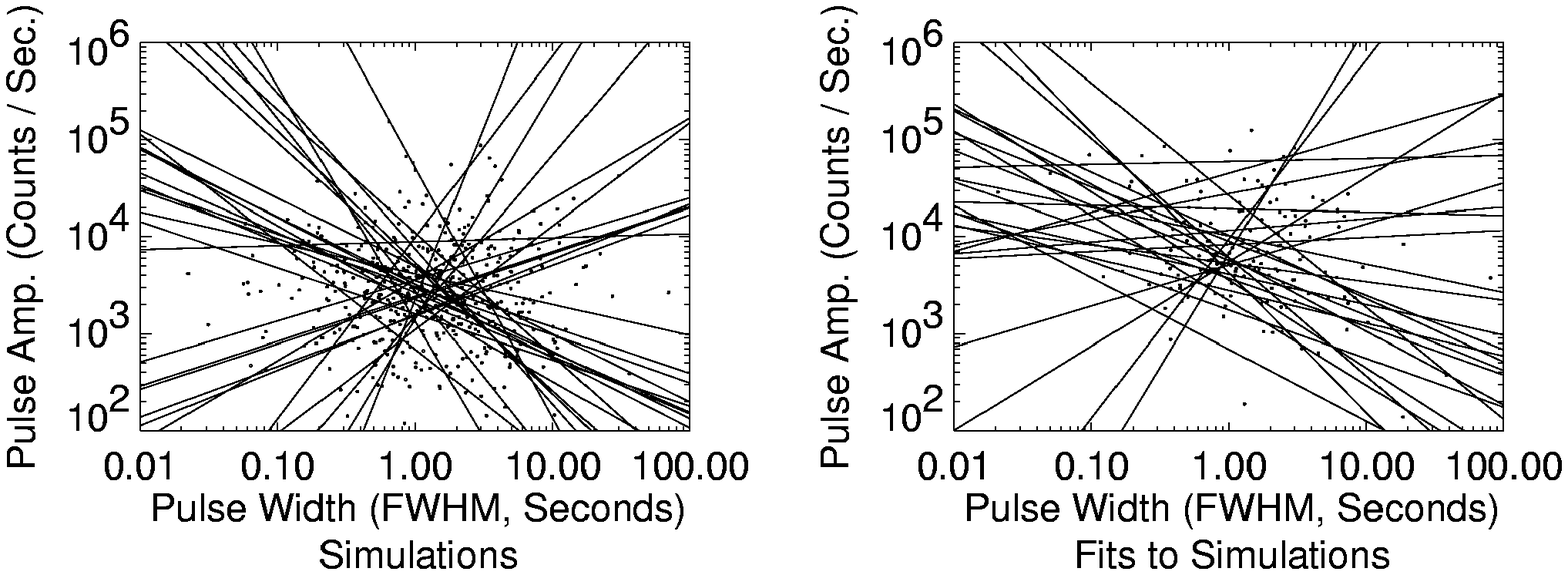}\caption{Pulse amplitudes versus pulse widths within bursts for bursts with strongest
correlations.  The lines show the fitted power law indices to pulses
in individual bursts with strong correlations.  Note that in the BATSE bursts (upper
panels), a large majority of the bursts show negative correlations (or
slopes) while in the simulations (lower panels), the numbers with
positive and negative correlations or slopes are much closer to equal.
\label{ampfwhm}}
\end{figure}

A more unambiguous test of the second type of correlation, intrinsic
correlations, can come from analysis of pulse widths and amplitudes of
pulses within bursts, because correlations between pulse
characteristics within bursts cannot be affected by the distances to
the sources, and are less likely to be affected by selection effects
due to the triggering process.  To this end, we have carried out
linear least squares fits to the logarithms of the pulse amplitudes
and widths in all actual BATSE bursts, and simulated bursts (before
and after fitting) which contain more than one pulse.  The results are
shown in Table~\ref{tab:ampfwhm}, which gives the numbers and
fractions of fits that show inverse correlations as determined from
the Spearman coefficients, and the probabilities that this would occur
by chance if there was no actual correlation, using the binomial
distribution.  It also gives the distributions of power-law indices
(slopes), which we denote as $\alpha$, in four bins: $\alpha < -1$,
$-1 < \alpha < 0$, $0 < \alpha < 1$, and $\alpha > 1$.  (For these
bins, the results are identical for three different linear regression
methods that are symmetric in the two variables being compared.  See
\cite{isobe:1990,lee:thesis}.)  The last column of
Table~\ref{tab:ampfwhm} gives the median power law index from the OLS
bisector method.  For all energy channels, a significant majority of
fits show inverse correlations between pulse widths and pulse
amplitudes \emph{within} bursts.  When we examine the actual BATSE
bursts for which the rank correlations have the greatest statistical
significance, shown in the upper panels of Figure~\ref{ampfwhm}, we
find that the vast majority of these show inverse correlations between
pulse widths and pulse amplitudes; in the bursts where the
correlations are positive, the correlations also tend to be less
statistically significant.  The pulse amplitudes most often vary as a
small negative power of the pulse width.  The power law indices are
significantly different from those relating pulse amplitude to pulse
width for the highest amplitude pulses in each
burst~\citep{petrosian:1999}.  \cite{ramirez-ruiz:1999b} found similar
results for the sample of 28 complex bursts fitted by
\cite{norris:1996}.  As noted by those authors, this anticorrelation
could be consistent with internal shock models of GRBs.

\begin{deluxetable}{crlrrrrr}
\tablecaption{Correlations Between Pulse Amplitude and Pulse Width Within Bursts, and the Distributions and Medians of the Fitted Power Law Index $\alpha$. \label{tab:ampfwhm}}
\tablehead{
\colhead{Energy} & \colhead{\% Neg.} & \colhead{Binom.} & \colhead{} & \colhead{} & \colhead{} & \colhead{} & \colhead{Med.} \\
\colhead{Channel} & \colhead{Corr.} & \colhead{Prob.} & \colhead{$\alpha < -1$} & \colhead{$-1 < \alpha < 0$} & \colhead{$0 < \alpha < 1$} & \colhead{$\alpha > 1$} & \colhead{$\alpha$}}
\startdata
1 & 65/94 = 69\% & 0.00020 & 21 & 42 & 22 & 9 & -0.37 \\
2 & 74/109 = 68\% & 0.00019 & 21 & 58 & 21 & 9 & -0.43 \\
3 & 82.5/116 = 71\% & \e{5.4}{-6} & 17 & 63 & 21 & 15 & -0.46 \\
4 & 26.5/35 = 76\% & 0.0023 & 3 & 26 & 5 & 1 & -0.55 \\
Sim. & 104.5/223 = 47\% & 0.35 & 39 & 55 & 65 & 64 & 0.55 \\
Fit to Sim. & 126.5/198 = 64\% & \e{9.3}{-5} & 32 & 86 & 53 & 24 & -0.39 \\
\enddata
\end{deluxetable}

Because of the possible far-reaching effects of this result, it is
important to ensure that this is not due to a selection or analysis
bias.  Our simulations can to some degree answer this question.
Table~\ref{tab:ampfwhm} also shows that there are no correlations
amplitude and pulse width within the simulated bursts.  In the fits to
the simulations, however, more bursts show a negative correlation
between pulse amplitude and pulse width than show a positive
correlation.  This asymmetry appears to be as large as it is for the
fits to actual BATSE data, which would suggest that the observed
tendency for higher amplitude pulses within bursts to be narrower
arises largely from a selection effect in the pulse-fitting procedure.
However, when we compare the fits to actual and simulated bursts for
which the rank correlations have the greatest statistical
significance, shown in the lower panels of Figure~\ref{ampfwhm}, we
find a different result.  In the simulated data, in the bursts with
correlations between pulse widths and pulse amplitudes with higher
statistical significance, the fraction that have positive correlations
between pulse widths and pulse amplitudes is similar to that in bursts
where the rank correlations have weaker statistical significance; the
asymmetry doesn't depend on the statistical significance of the
correlations.  This is unlike the fits to actual bursts, where almost
all of the bursts with the most statistically significant correlations
show a negative slope~\citep{petrosian:1999}.  Therefore, the observed
inverse correlations between pulse widths and pulse amplitudes within
actual bursts appear to arise in part from intrinsic properties of the
sources.

\begin{figure}\plotone{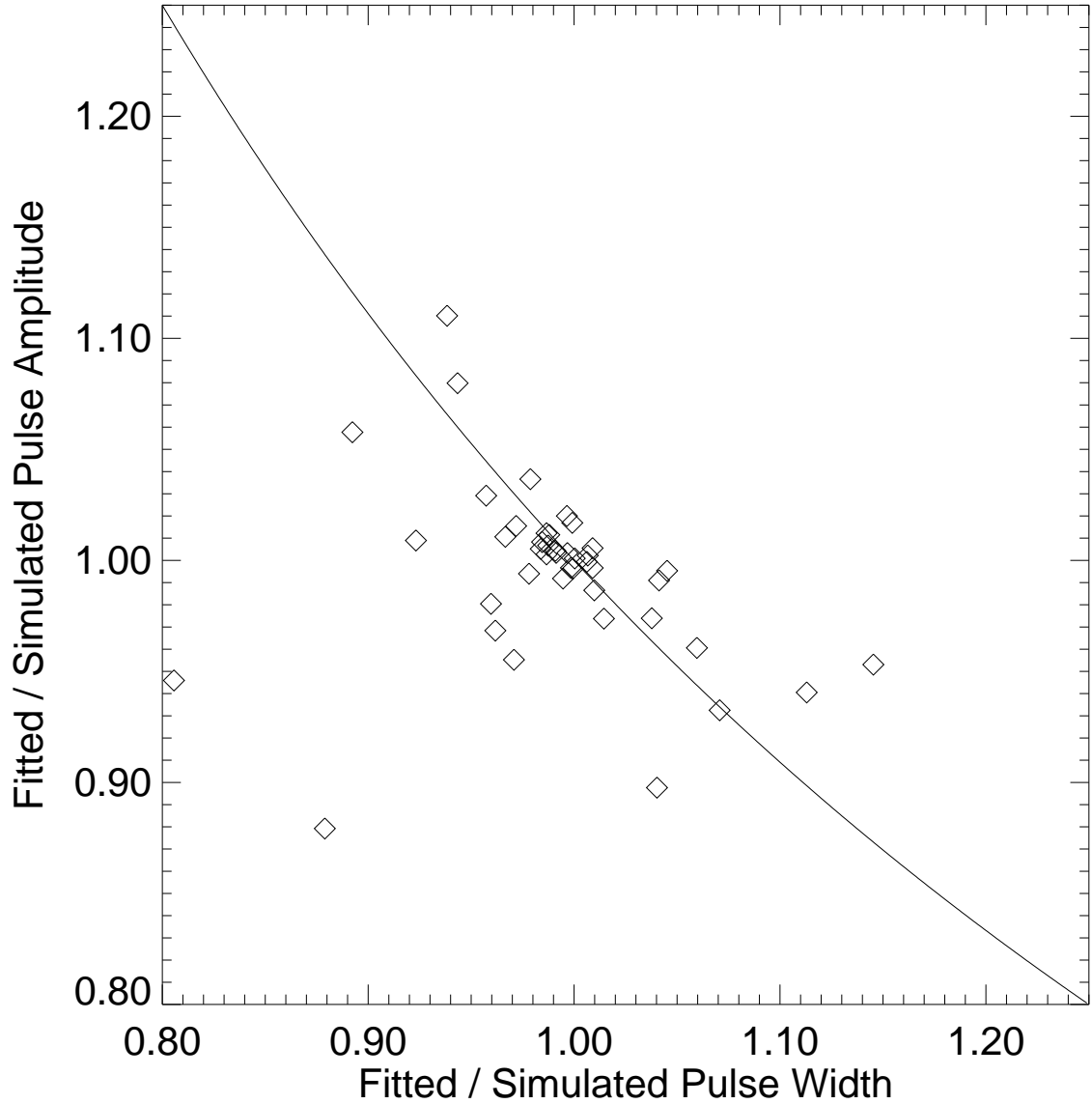}\caption{Ratios of fitted to simulated pulse amplitudes versus ratios of fitted to simulated pulse widths, with line of constant count fluence, for single-pulse simulated bursts.  Note that the errors arising from the fitting procedure for these
quantities are anticorrelated, which would cause a bias in the fitting
procedure favoring anticorrelated pulse amplitudes and pulse widths.  \label{sim_rampvsrw}}
\end{figure}

However, some caution is necessary in the interpretation of these
results.  This is because we find correlations between the errors in
the fitted pulse parameters by comparing the parameters used in the
simulations with those obtained from the fits to the simulations.  For
simulated bursts consisting of a single pulse in both the original
simulation and in the fit, the identification of pulses between the
simulation and the fit is unambiguous and unaffected by the effects of
missing pulses.  Figure~\ref{sim_rampvsrw} shows that the errors in
the fitted pulse amplitudes and the fitted pulse widths tend to have
an inverse correlation; when the fitted amplitude is larger than the
original amplitude, the fitted width tends to be smaller than the
original width, and vice versa.  The same effect also appears when we
compare the highest amplitude pulses from all bursts, or all pulses
matched between the simulations and the fits to the simulations.  This
selection effect may cause weak inverse correlations between pulse
amplitude and pulse width within fits to actual or simulated bursts,
so it may be another reason why a large majority of both actual BATSE
bursts and fits to simulated bursts show an inverse correlation
between pulse amplitude and pulse width within the bursts, as found
here and by \cite{ramirez-ruiz:1999b}.  \emph{However, we conclude
that the evidence for intrinsic correlation between pulse amplitude
and width is weak and requires further study.  Therefore, caution
should be exercised in the interpretation of this result, in
particular in using it as evidence against the external shock model.}

\subsection{Other Timescales}

Cosmological time dilation must affect all timescales within bursts,
not only pulse widths.  Some of these timescales may provide a more
robust test of cosmological time dilation.  This is because use of a
pulse width as a burst duration is subject to the following
uncertainty.  Because of the spectral shift due to cosmological
redshift, for the dimmer, hence more distant, bursts, BATSE will be
detecting higher energy rest frame photons.  gamma-rays were
originally produced at higher energies but had redshifted to lower
energies when they were detected.  Since both burst durations
\citep{fenimore:1995} and pulses \citep{lee:2000} tend to be shorter
at higher energies, this would weaken the correlations between
amplitude and width due to time dilation.

\begin{figure}
\epsfig{file=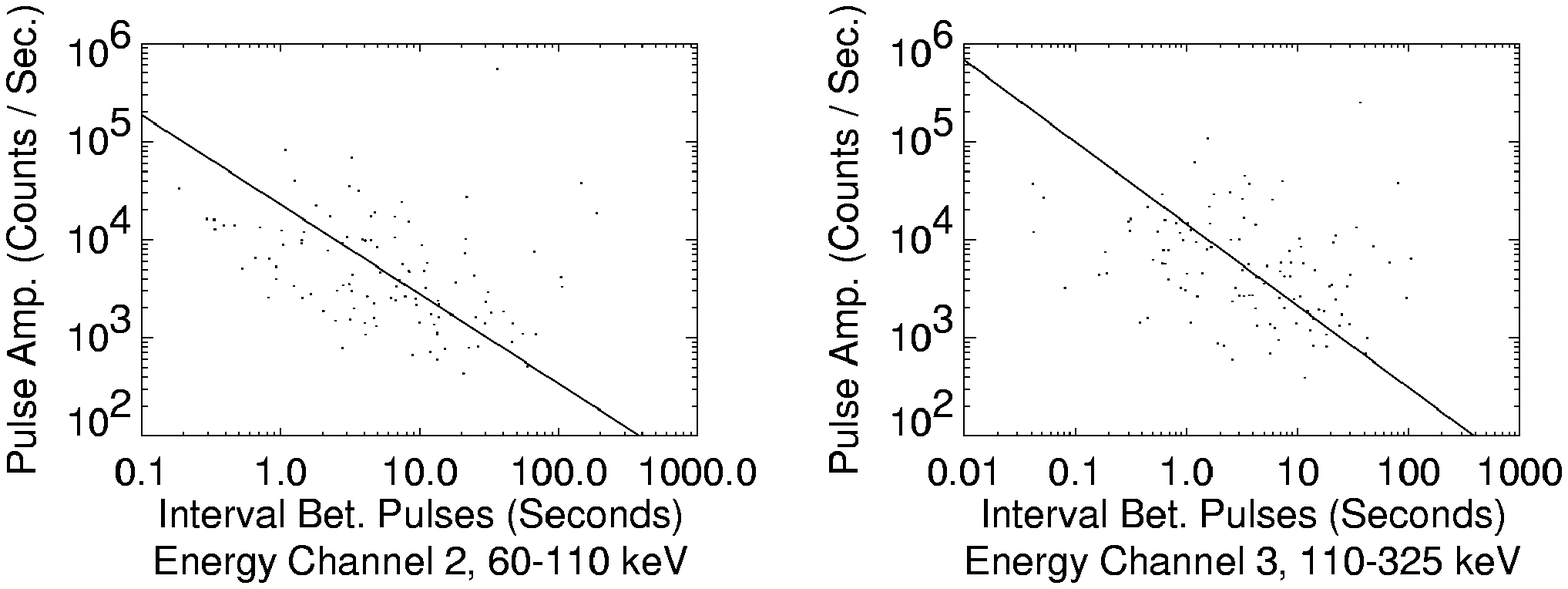}
\epsfig{file=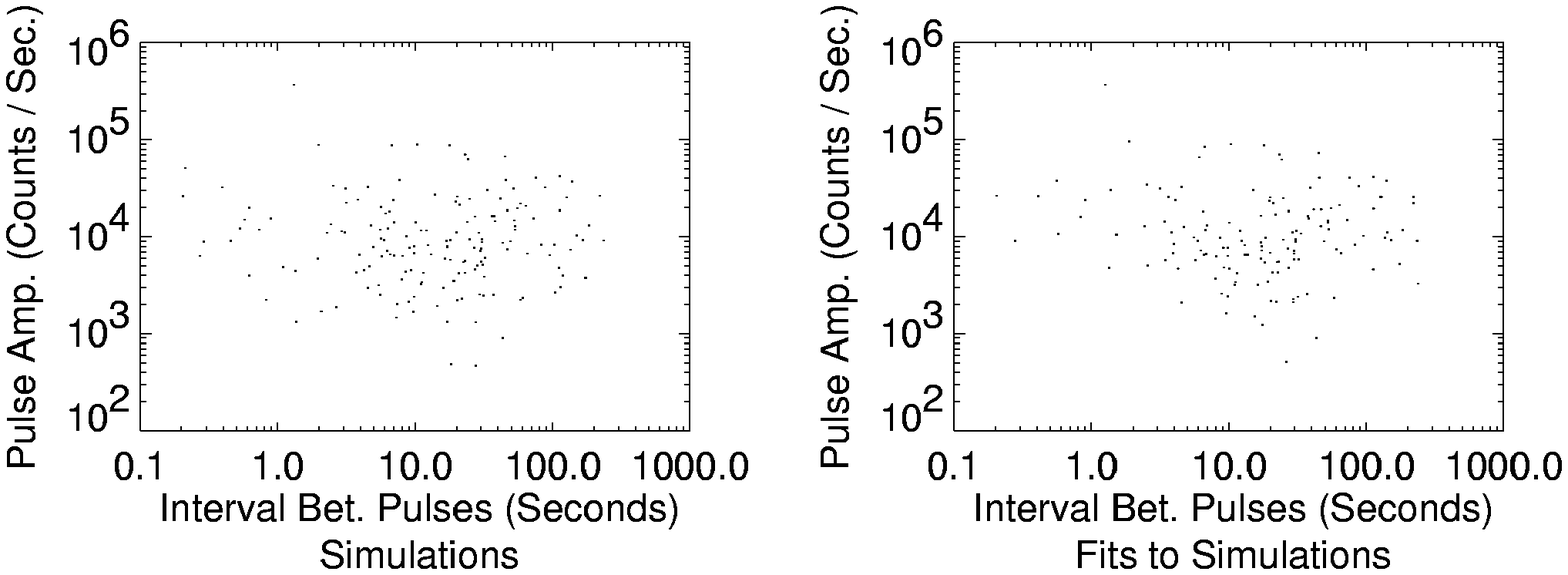}\caption{Highest pulse amplitude versus interval between two highest amplitude pulses in each burst.  In the initial simulations and the fits to the simulations (bottom panels), the correlations were insignificant, so no fits were made.  \label{peakampvsint}}
\end{figure}

We have seen earlier that the \emph{intervals between the peak times} of the
two highest amplitude pulses in each burst do not appear to increase
or decrease with energy, so that cosmological redshift of photon energies
should not affect these intervals.  As shown in
the upper panels of Figure~\ref{peakampvsint} and columns~(b) of
Table~\ref{tab:peak_ampvswint}, these intervals also show a
significant inverse correlation with the amplitudes of the highest
amplitude pulses in the actual bursts, so they are shorter for
brighter bursts.

Such a trend does not seem to be present in the fits to the simulated
data, and is not present in the initial simulated data by design (See
Figure~\ref{peakampvsint}, lower panels, and bottom two rows of
Table~\ref{tab:peak_ampvswint}, columns~(b).)  The distributions are
very similar for the simulated bursts and for the fits to the
simulations, although the fits to simulations tend to miss points when
both the peak amplitudes and the intervals between the two highest
amplitude pulses are small.  Therefore, it appears that the
correlations observed in the fits to actual bursts observed by BATSE
are not caused by selection effects in the fitting procedure, but may
arise from cosmological time dilation or from intrinsic properties of
the bursts.  An early study of time dilation using the intervals
between pulses found inconsistent results~\citep{neubauer:1996}, but a
number for later studies have found evidence of time
dilation~\citep{norris:1996b,deng:1998,deng:1998b} consistent with our
results.

To see if some kind of correlation is present among pulses within
bursts, we compare pulse amplitudes with time intervals between pulses
within bursts as follows: For each burst time profile consisting of
three or more pulses, we order the individual pulses by decreasing
pulse amplitude.  Then we look for correlations between the amplitude
of each pulse and the absolute value of the intervals between it and
the pulse with the next lower amplitude.  The results are shown in
Table~\ref{tab:ampint}.  There appears to be a more frequent
occurrence of inverse correlations than positive correlations between
pulse amplitudes and intervals between pulses within bursts in the
BATSE data, but this is statistically insignificant in all energy
channels except possibly channel~1.  This table also shows that the
fitting procedure does not introduce any significant bias.

\begin{deluxetable}{crlrrrrr}
\tablecaption{Correlations Between Pulse Amplitude and Intervals Between Pulses Within Bursts, and Distributions and Medians of the Fitted Power Law Index $\alpha$. \label{tab:ampint}}
\tablehead{
\colhead{Energy} & \colhead{\% Neg.} & \colhead{Binom.} & \colhead{} & \colhead{} & \colhead{} & \colhead{} & \colhead{Med.} \\
\colhead{Channel} & \colhead{Corr.} & \colhead{Prob.} & \colhead{$\alpha < -1$} & \colhead{$-1 < \alpha < 0$} & \colhead{$0 < \alpha < 1$} & \colhead{$\alpha > 1$} & \colhead{$\alpha$}}
\startdata
1 & 42/62 = 68\% & 0.0052 & 5 & 38 & 14 & 5 & -0.47 \\
2 & 54/89 = 61\% & 0.044 & 9 & 49 & 26 & 5 & -0.44 \\
3 & 55/95 = 58\% & 0.12 & 4 & 51 & 34 & 6 & -0.32 \\
4 & 17.5/24 = 73\% & 0.064 & 2 & 15 & 5 & 2 & -0.48 \\
Sim. & 44/156 = 51\% & 0.11 & 23 & 43 & 52 & 38 & 0.67 \\
Fit to Sim. & 74/132 = 56\% & 0.16 & 17 & 60 & 34 & 21 & -0.29 \\
\enddata
\end{deluxetable}

Finally, it should also be noted that the fitted power law indices for
highest pulse amplitude versus width of the highest amplitude pulse
are smaller than -1, which is inconsistent with purely cosmological
effects.  For a given variation in the highest pulse amplitude, the
corresponding variation in pulse width is too great to be accounted
for by only cosmological time dilation.  We have also seen that within
individual bursts, higher amplitude pulses have a strong tendency to
be narrower, which must result from intrinsic properties of the GRB
sources themselves.  It seems likely that the observed correlation
between the highest pulse amplitude and the width of the highest
pulses in each burst could result from a combination of cosmological
and non-cosmological effects.

One of the possible intrinsic effects that could contribute to the
inverse correlations of pulse widths with pulse amplitudes is that the
total energy in a burst, or within individual pulses, might tend to
fall within a limited range, or might have an upper limit.  This would
be the case if, for example, the fluence of a burst were a better
measure of distance than the peak flux.  In the next section, we
repeat the above tests using the fluence instead of peak flux as a
measure of the strengths of bursts and pulses.

On the other hand, the power law indices for highest pulse amplitude
versus the time interval between the peaks of the two highest pulses
in each burst may be consistent with the expected results of
cosmological time dilation alone.  Furthermore, it seems likely that
this correlation is less affected by intrinsic properties of bursters
or by selection effects than the correlation between the highest pulse
amplitude and the width of the same pulse in each burst.  For example,
if the range of radiated energy in entire bursts or in individual
pulses, were limited by the production mechanism, or by selection
effects, this would be far less likely to affect intervals between
pulses than to affect pulse widths.

\subsection{Integrated Luminosity as a Standard Candle}

\cite{petrosian:1996} have suggested that the integrated luminosities
of bursts, measured using either energies or photons, are likely to be
better standard candles than their peak luminosities.  This would be
the case if the total energy output of bursters fall in a narrow range
of values, and much of the variation in flux results from the broad
range of burst durations.  \cite{petrosian:1996b,lee:1997} have also
found that the energy fluences of bursts and their durations show a
positive correlation, which is the opposite of what cosmological time
dilation should cause.  In what follows we carry out similar tests for
bursts and for pulses within individual bursts.  We shall see that the
count fluences of bursts and pulse widths show a positive correlation,
while the count fluences of bursts and time intervals between pulses
show no correlation, and neither of these effects can arise from
cosmological effects.  However, determining the significance of some
of these correlations is difficult because the simulated bursts were
generated with no correlations between pulse width and pulse
amplitude, and therefore have a positive correlation between pulse
width and pulse count fluence.

\begin{figure}
\epsfig{file=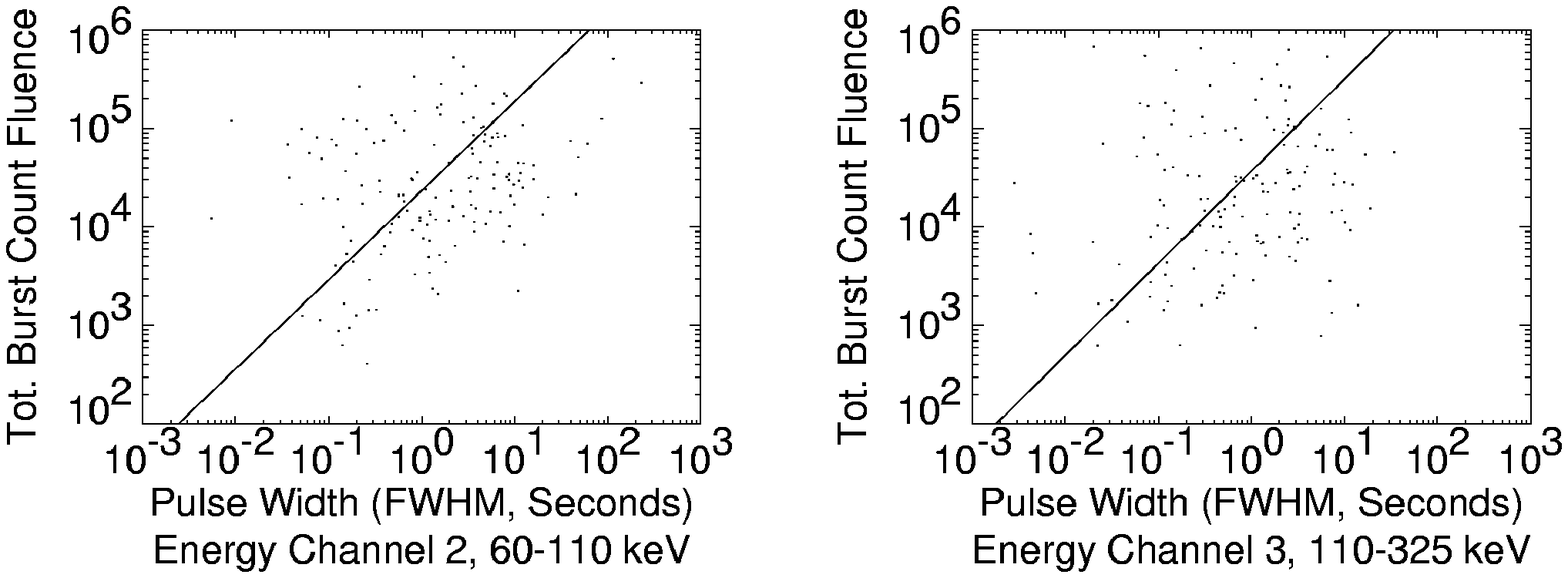}
\epsfig{file=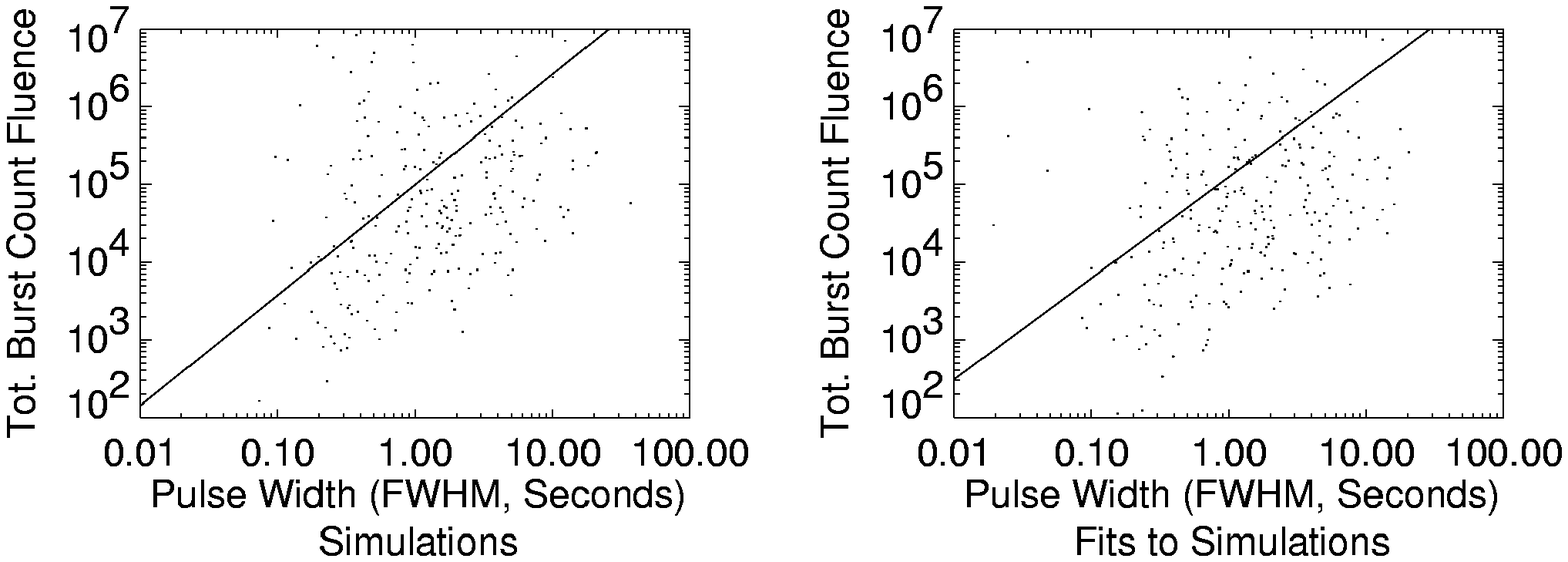}\caption{Total count fluence versus pulse width (FWHM) of highest amplitude pulse in each burst. \label{tareavsw}}
\end{figure}

In Figure~\ref{tareavsw}, we show that the pulse widths of the highest
amplitude pulses have positive correlations with the total count
fluences of each fit that appear to be significant in all energy
channels except perhaps in channel~3.  (See also
Table~\ref{tab:tareavswint}, columns~(a).)  The positive correlation
appears somewhat stronger in the fits to simulations than in the
simulations.  As mentioned above, this makes the interpretation of
this result difficult.

\begin{deluxetable}{crlrrlr}
\tablecaption{Correlation Between Total Count Fluence and (a) Pulse Width (FWHM) of Highest Amplitude Pulse in Each Burst and (b) Interval Between Two Highest Pulses in Each Burst, and the Fitted Power Law Index $\beta$. \label{tab:tareavswint}}
\tablehead{
\colhead{Energy} & \multicolumn{3}{c}{(a) Pulse Width} & \multicolumn{3}{c}{(b) Interval} \\
\colhead{Channel} & \colhead{$r_{s}$} & \colhead{Prob.} & \colhead{$\alpha$} & \colhead{$r_{s}$} & \colhead{Prob.} & \colhead{$\alpha$}}
\startdata
1 & 0.29 & \e{2.4}{-4} & $0.89 \pm 0.06$ & 0.096 & 0.36 & $0.99 \pm 0.03$ \\
2 & 0.27 & \e{2.7}{-4} & $0.91 \pm 0.05$ & 0.15 & 0.11 & $1.03 \pm 0.06$ \\
3 & 0.17 & 0.023 & $0.93 \pm 0.04$ & 0.26 & \e{4.5}{-3} & $0.98 \pm 0.06$ \\
4 & 0.33 & \e{5.8}{-3} & $0.95 \pm 0.05$ & 0.25 & 0.15 & $1.09 \pm 0.10$ \\
Simulation & 0.23 & \e{8.9}{-5} & $1.42 \pm 0.18$ & -0.04 & 0.59 & \nodata \\
Fit to Sim. & 0.33 & \e{1.3}{-8} & $1.30 \pm 0.17$ & -0.06 & 0.51 & \nodata \\
\enddata
\end{deluxetable}

\begin{figure}
\epsfig{file=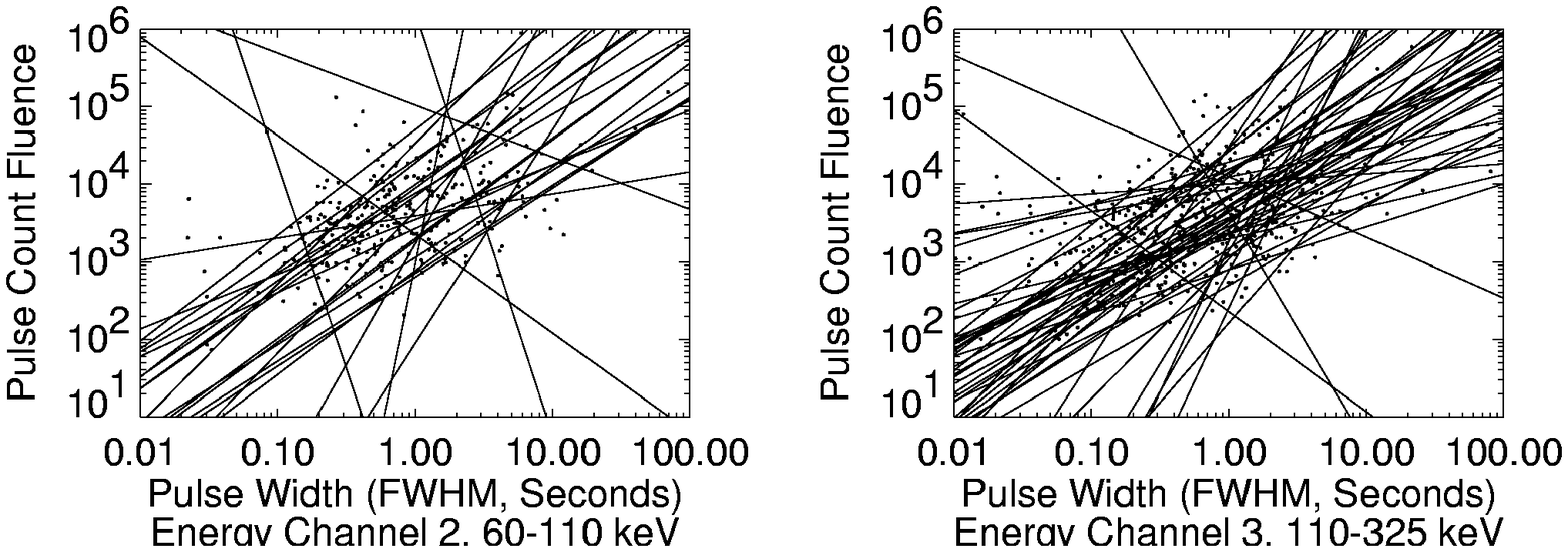}
\epsfig{file=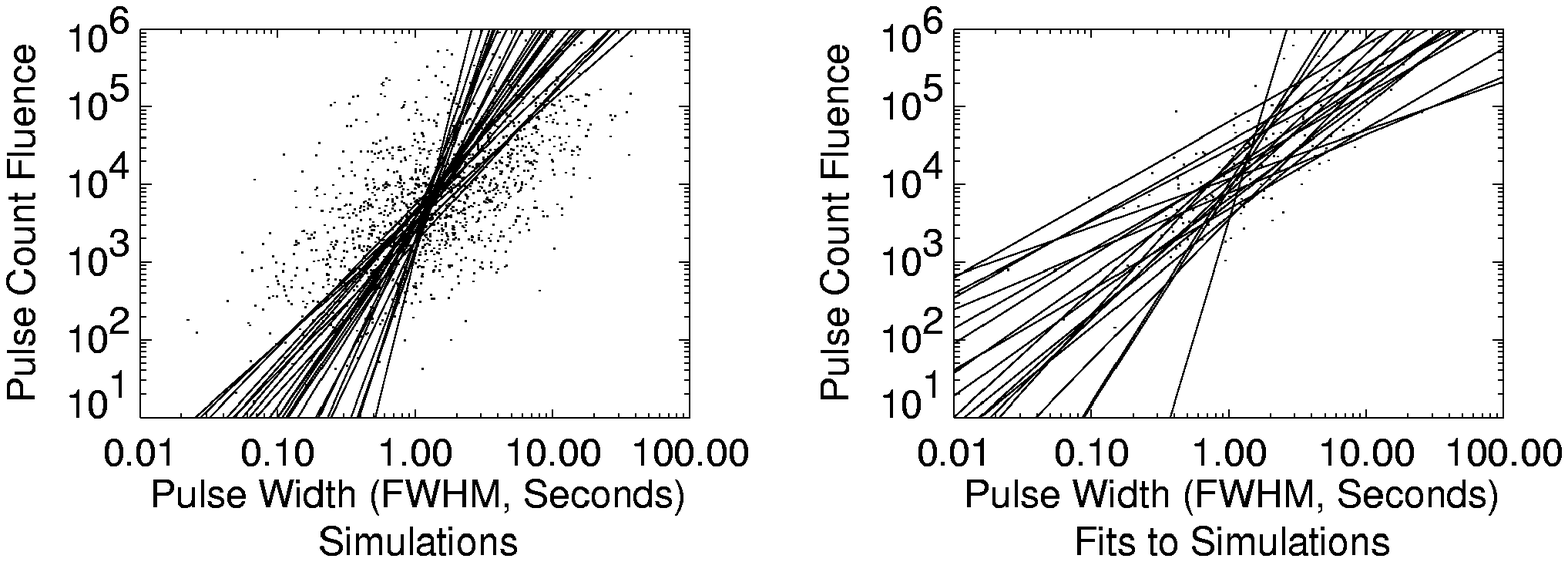}\caption{Pulse count fluences versus pulse widths within bursts for bursts with strongest correlations.  The lines show the fitted power law indices to pulses in individual bursts with strong correlations.  In the BATSE bursts
(upper panels), nearly all bursts show positive correlations (or
slopes), indicating that the distributions of pulse amplitudes within
the individual bursts are narrow.  In the simulated bursts (lower
panels), all bursts show positive correlations between pulse amplitude
and pulse width because of the design of the
simulation. \label{areafwhm}}
\end{figure}

Correlations between pulse width and pulse count fluence \emph{within}
bursts do not appear to have been studied before.  In
Table~\ref{tab:areafwhm}, we show the distribution and some moments of
the power law index $\beta$, which is obtained from linear fits to the
logarithms of the fluence and widths of pulses in individual bursts.
As evident, a significant majority of fits in all energy channels and
in the simulations show strong positive correlations between pulse
width and pulse count fluence within individual bursts.  (See
Table~\ref{tab:areafwhm} and the upper panels of
Figure~\ref{areafwhm}.)  Pulse count fluences most often vary as a
large positive power of the pulse width.  (Because more bursts have
$|\beta| > 1$ than $|\beta| < 1$, taking the median of the reciprocal
of $\beta$ is more appropriate.)

\begin{deluxetable}{crlrrrrr}
\tablecaption{Correlations Between Pulse Count Fluence and Pulse Width Within Bursts, and Distributions and Medians of the Fitted Power Law Index $\beta$. \label{tab:areafwhm}}
\tablehead{
\colhead{Energy} & \colhead{\% Pos.} & \colhead{Binom.} & \colhead{} & \colhead{} & \colhead{} & \colhead{} & \colhead{} \\
\colhead{Channel} & \colhead{Corr.} & \colhead{Prob.} & \colhead{$\beta < -1$} & \colhead{$-1 < \beta < 0$} & \colhead{$0 < \beta < 1$} & \colhead{$\beta > 1$} & \colhead{$\displaystyle\frac{1}{{\text{Med.}}(1 / \beta)}$}}
\startdata
1 & 66/94 = 70\% & \e{8.9}{-5} & 14 & 15 & 23 & 42 & 1.88 \\
2 & 77.5/109 = 71\% & \e{1.0}{-5} & 20 & 13 & 25 & 51 & 1.46 \\
3 & 90.5/116 = 78\% & $<10^{-16}$ & 16 & 12 & 38 & 50 & 1.29 \\
4 & 27/35 = 77\% & 0.0013 & 3 & 5 & 17 & 10 & 1.03 \\
Sim. & 198/223 = 89\% & $<10^{-16}$ & 14 & 7 & 37 & 165 & 1.59 \\
Fit to Sim. & 167/198 = 84\% & $<10^{-16}$ & 25 & 12 & 54 & 103 & 1.41 \\
\enddata
\end{deluxetable}

The last two rows of Table~\ref{tab:areafwhm} and the lower panels of
Figure~\ref{areafwhm} show that the correlations in the fits to
simulations are similar, though somewhat weaker in the original
simulations, so that the observed correlation for the BATSE bursts is
probably not a result of the fitting procedure.

\begin{figure}\plotone{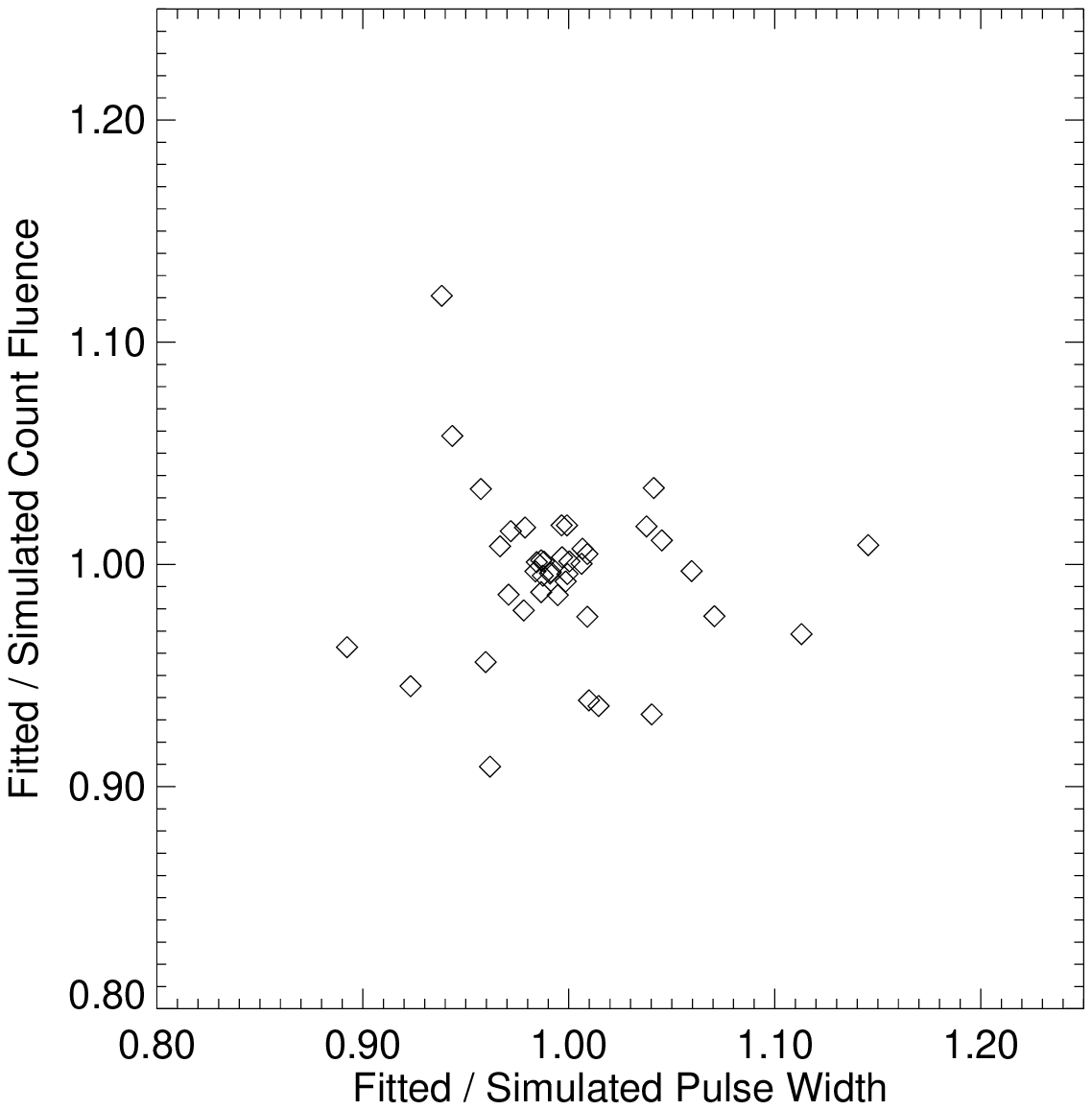}\caption{Ratios of fitted to simulated pulse count fluences versus ratios of fitted to simulated pulse widths for single-pulse simulated bursts.  Note that unlike
Figure~\ref{sim_rampvsrw}, the errors here do not show any significant
correlation.  \label{sim_rareavsrw}}
\end{figure}

Figure~\ref{sim_rareavsrw} shows that there are no significant
correlations between the errors in the fitted count fluences and the
fitted pulse widths for simulated bursts consisting of a single pulse
in both the simulation and the fit.  Therefore, the uncorrelated
errors in the pulse count fluences and pulse widths would tend to
smear out any existing correlations rather than to create
correlations, which is what we have seen above.

\begin{figure}
\epsfig{file=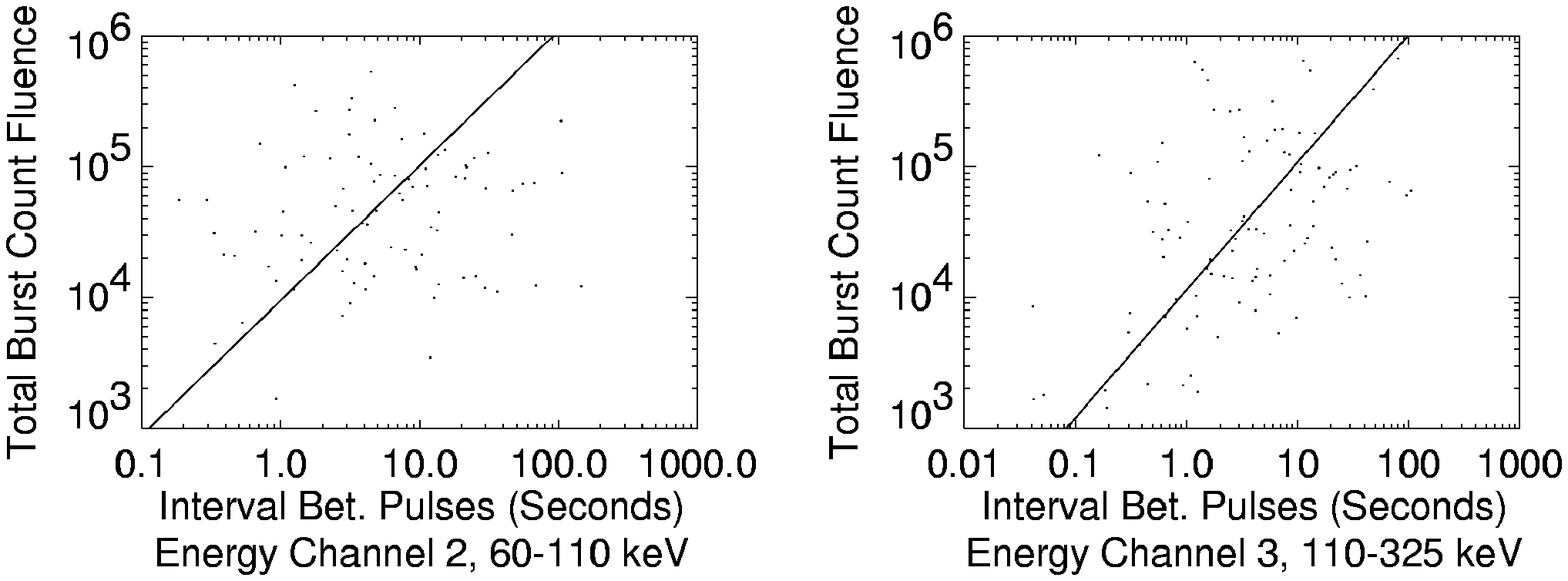}
\epsfig{file=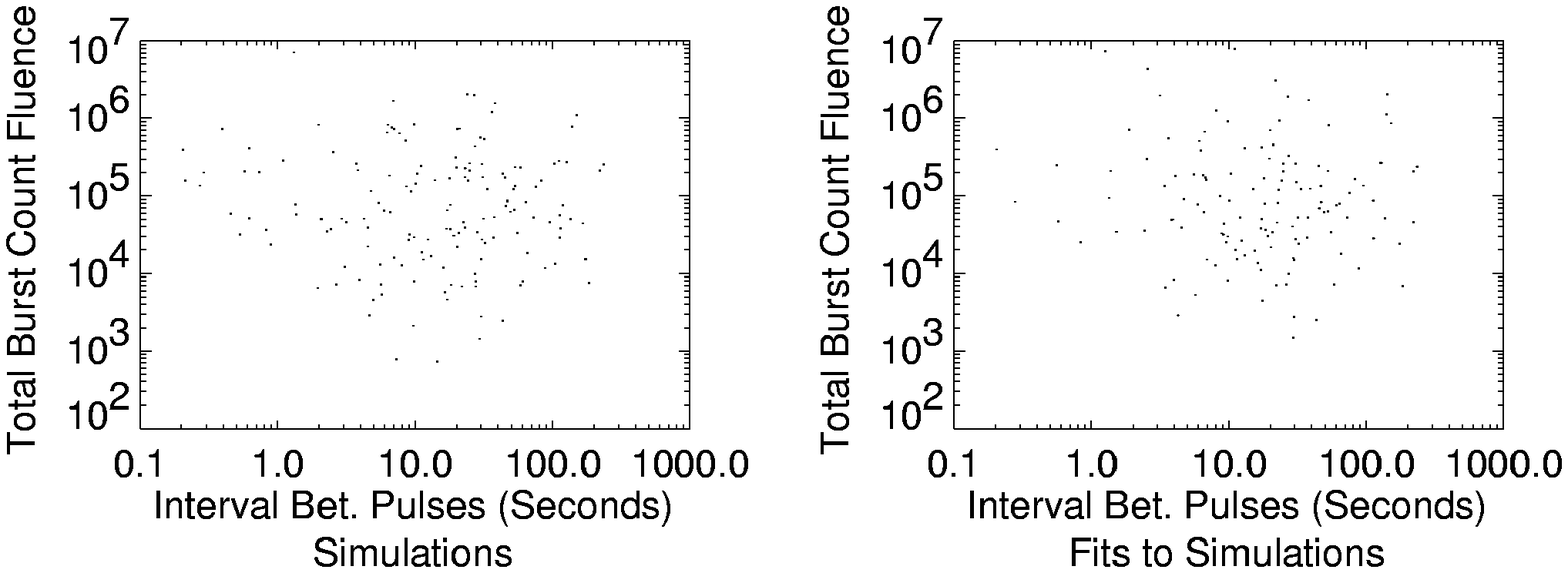}\caption{Total count fluence versus interval between two highest pulses in each burst.  In the initial simulations and the fits to the simulations (bottom panels), the correlations were insignificant, so no fits were made. \label{tareavsint}}
\end{figure}

The relation between the total count fluence and time interval between
the two highest amplitude pulses in the actual and simulated bursts
are shown in Figure~\ref{tareavsint} and columns~(b) of
Table~\ref{tab:tareavswint}.)  The two quantities have positive
correlations in all energy channels in the actual BATSE bursts, as
determined from the Spearman rank-order correlation coefficients.
However, the correlation is statistically insignificant in all
channels, except perhaps in channel~3.

The distributions of the total burst count fluence versus \emph{the
intervals between the peak times} of the two highest amplitude pulses
in each burst are very similar for the simulated bursts and for the
fits to the simulations, although the fits to simulations tend to miss
points when the intervals between the two highest amplitude pulses are
small.  However, columns~(b) of Table~\ref{tab:tareavswint} show no
significant correlation for either the simulations or the fits to the
simulations.  This indicates that any correlation that may be present
in the BATSE bursts is intrinsic to the radiative process.

We can also compare the count fluences of individual pulses with the
time intervals between pulses within bursts.  The results, shown in
Table~\ref{tab:areaint}, show no statistically significant
correlations between these two quantities.  The simulations and fits
to simulations also show no statistically significant correlations.

\begin{deluxetable}{crlrrrrr}
\tablecaption{Correlations Between Pulse Count Fluence and Intervals Between Pulses Within Bursts, and Distributions and Medians of the Fitted Power Law Index $\beta$. \label{tab:areaint}}
\tablehead{
\colhead{Energy} & \colhead{\% Pos.} & \colhead{Binom.} & \colhead{} & \colhead{} & \colhead{} & \colhead{} & \colhead{} \\
\colhead{Channel} & \colhead{Corr.} & \colhead{Prob.} & \colhead{$\beta < -1$} & \colhead{$-1 < \beta < 0$} & \colhead{$0 < \beta < 1$} & \colhead{$\beta > 1$} & \colhead{$\displaystyle\frac{1}{{\text{Med.}}(1 / \beta)}$}}
\startdata
1 & 33/62 = 53\% & 0.61 & 21 & 8.5 & 14.5 & 18 & 13.5 \\
2 & 49.5/89 = 56\% & 0.29 & 19 & 18 & 18 & 34 & 6.5 \\
3 & 54.5/95 = 57\% & 0.15 & 17 & 20 & 24 & 34 & 3.1 \\
4 & 12/24 = 50\% & 1.0 & 4 & 9 & 7 & 4 & -4.6 \\
Sim. & 71.5/156 = 46\% & 0.30 & 57 & 13 & 18 & 68 & 4.1 \\
Fit to Sim. & 63/132 = 48\% & 0.60 & 45 & 18 & 17 & 52 & 34 \\
\enddata
\end{deluxetable}

In summary, all correlations between \emph{pulse} count fluences and
pulse widths are positive, and probably result from the simple fact
that pulses of longer duration tend to contain more counts.  The
correlation between total \emph{burst} count fluence and the width of
the highest amplitude pulse in each burst is probably a result of this
correlation and the fact that the majority of the total count fluence
of a burst is often contained in a single pulse.  The cosmological
effects have been overwhelmed by other effects.

It is not clear why there appears to be no correlation between total
burst count fluence and the interval between the two highest pulses in
each burst.  One possibility is that most of the observed bursts are
sufficiently far away that the count fluence varies very little with
luminosity distance.  However, this would place many bursts at
redshifts of $z > 10$, which seems unlikely given current evidence.

\section{Other Correlations}
\label{sec:othercorr}

\subsection{Correlations Between Flux and Fluence}

Since the count fluence of a pulse scales as the product of its
amplitude and its width, and a factor involving the peakedness $\nu$,
or equivalently, since the amplitude of a pulse scales as its count
fluence divided by its width, again with a factor involving $\nu$,
various selection effects could cause observed pulse amplitudes and
widths to have an inverse correlation or cause observed pulse count
fluences and widths to have a positive correlation.

\begin{figure}
\epsfig{file=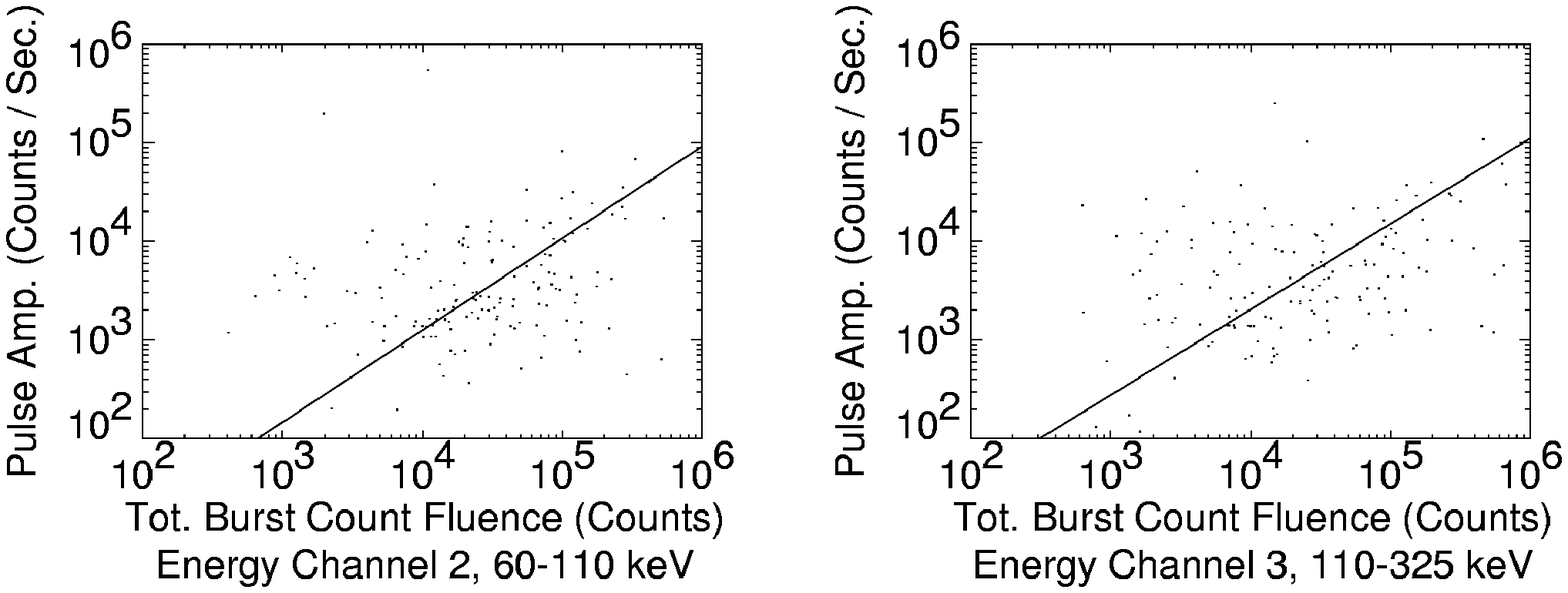}
\epsfig{file=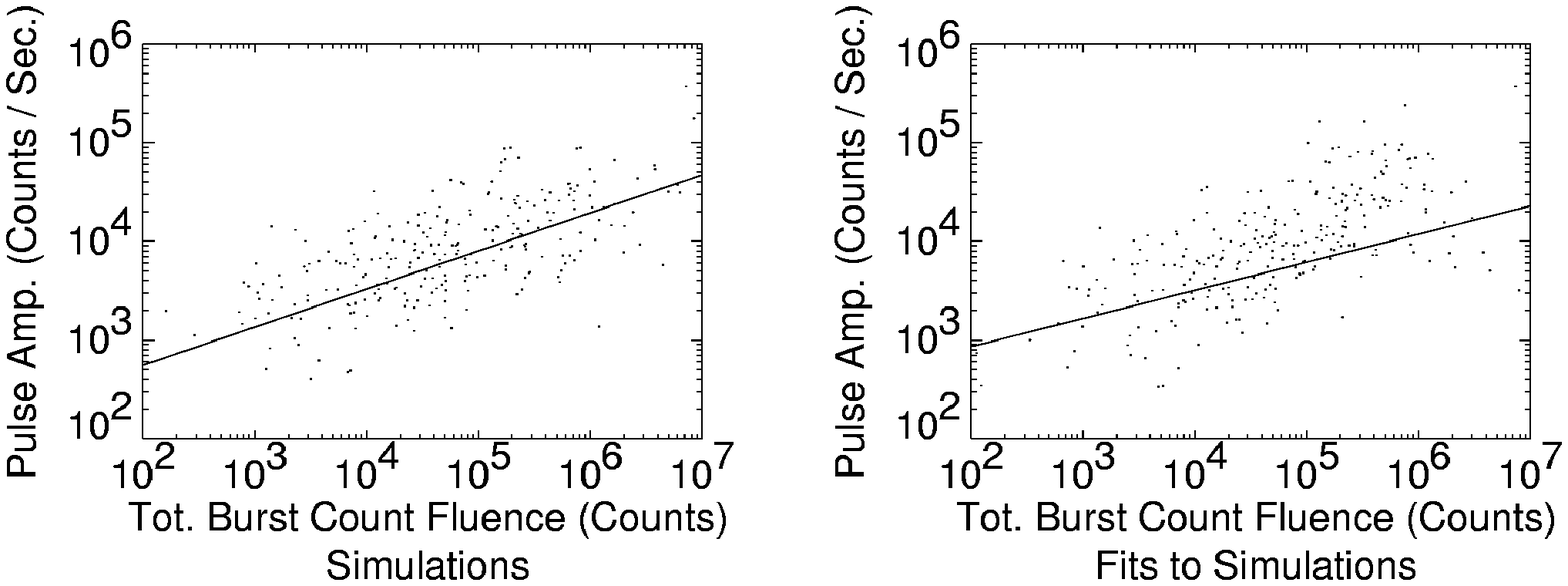}\caption{Amplitude of highest amplitude pulse versus total count fluence in each burst. \label{peak_ampvstarea}}
\end{figure}

Figure~\ref{peak_ampvstarea} and Table~\ref{tab:peak_ampvstarea} show
that there are no strong correlations between the amplitudes of the
highest amplitudes pulses and the total count fluences of the BATSE
bursts, in any energy channel.  This result is somewhat unexpected,
because even in the absence of cosmological effects, we would expect
both peak flux and total fluence to scale approximately as the inverse
square of the luminosity distance to the sources (the effects of the
time dilation factor $1 + z$ are much smaller), and hence to have a
positive correlation with each other.  The results from our
simulations are not helpful because the simulated bursts were also
generated with a strong positive correlation between pulse amplitude
and pulse count fluence.  It appears that selection effects in the
pulse-fitting procedure tend to weaken these positive correlations,
shown in the last two rows of Table~\ref{tab:peak_ampvstarea}, when we
compare the simulations with the fits to the simulations.  The absence
of correlation in the actual bursts may indicate that the intrinsic
range of the \emph{effective durations}, \emph{i.e.}  the total
fluences divided by the peak fluxes \citep{lee:1997}, is large enough
to smear out distance effects expected in the distribution of fluences
and peak fluxes.  It also suggests that if one of the two brightness
measures is a good indicator of distance, then the other cannot be,
probably due to selection effects, or due to cosmological evolution of
the sources.

\clearpage

\begin{deluxetable}{crlr}
\tablecaption{Correlation Between Amplitude of Highest Amplitude Pulse
and Total Count Fluence in Each Burst, and the Fitted Power Law Index
$\gamma$. \label{tab:peak_ampvstarea}}
\tablehead{
\colhead{Energy
Channel} & \colhead{$r_{s}$} & \colhead{Prob.} & \colhead{$\gamma$}}
\startdata
1 & 0.13 & 0.096 & $0.92 \pm 0.07$ \\
2 & 0.043 & 0.57 & $0.93 \pm 0.05$ \\
3 & 0.15 & 0.053 & $0.87 \pm 0.04$ \\
4 & -0.036 & 0.77 & $0.96 \pm 0.12$ \\
Simulation & 0.65 & \e{4.4}{-36} & $0.39 \pm 0.03$ \\
Fit to Sim. & 0.61 & \e{8.9}{-31} & $0.28 \pm 0.06$ \\
\enddata
\end{deluxetable}

However, when we consider the relation for pulses within individual
bursts, we find that a significant majority of bursts in all energy
channels show a positive correlation between pulse count fluence and
amplitude within bursts.  (See Table~\ref{tab:amparea}.) In every
energy channel, the majority of bursts have pulse amplitudes varying
as a small positive power $\gamma$ of the pulse count fluence within
bursts.  Most simulated bursts, as expected, show a positive
correlation between pulse amplitude and pulse count fluence, but in
the fits to the simulations, fewer bursts show a positive correlation.
Therefore, the actual correlation in the BATSE bursts may have been
weakened by selection effects in the pulse-fitting procedure.

\begin{deluxetable}{crlrrrrr}
\tablecaption{Correlations Between Pulse Amplitude and Pulse Count Fluence Within Bursts, and the Distributions and Medians of the Fitted Power Law Index $\gamma$. \label{tab:amparea}}
\tablehead{
\colhead{Energy} & \colhead{\% Pos.} & \colhead{Binom.} & \colhead{} & \colhead{} & \colhead{} & \colhead{} & \colhead{Med.} \\
\colhead{Channel} & \colhead{Corr.} & \colhead{Prob.} & \colhead{$\gamma < -1$} & \colhead{$-1 < \gamma < 0$} & \colhead{$0 < \gamma < 1$} & \colhead{$\gamma > 1$} & \colhead{$\gamma$}}
\startdata
1 & 71/94 = 76\% & \e{7.2}{-7} & 8 & 18 & 54 & 14 & 0.48 \\
2 & 86.5/109 = 79\% & $<10^{-16}$ & 3 & 23 & 62 & 21 & 0.61 \\
3 & 85/116 = 73\% & \e{4.8}{-7} & 6 & 22 & 75 & 13 & 0.63 \\
4 & 24/35 = 69\% & 0.028 & 3 & 8 & 19 & 5 & 0.61 \\
Sim. & 185/223 = 83\% & $<10^{-16}$ & 11 & 24 & 172 & 16 & 0.34 \\
Fit to Sim. & 142.5/198 = 72\% & $<10^{-16}$ & 15 & 40 & 121 & 18 & 0.20 \\
\enddata
\end{deluxetable}

\begin{figure}\plotone{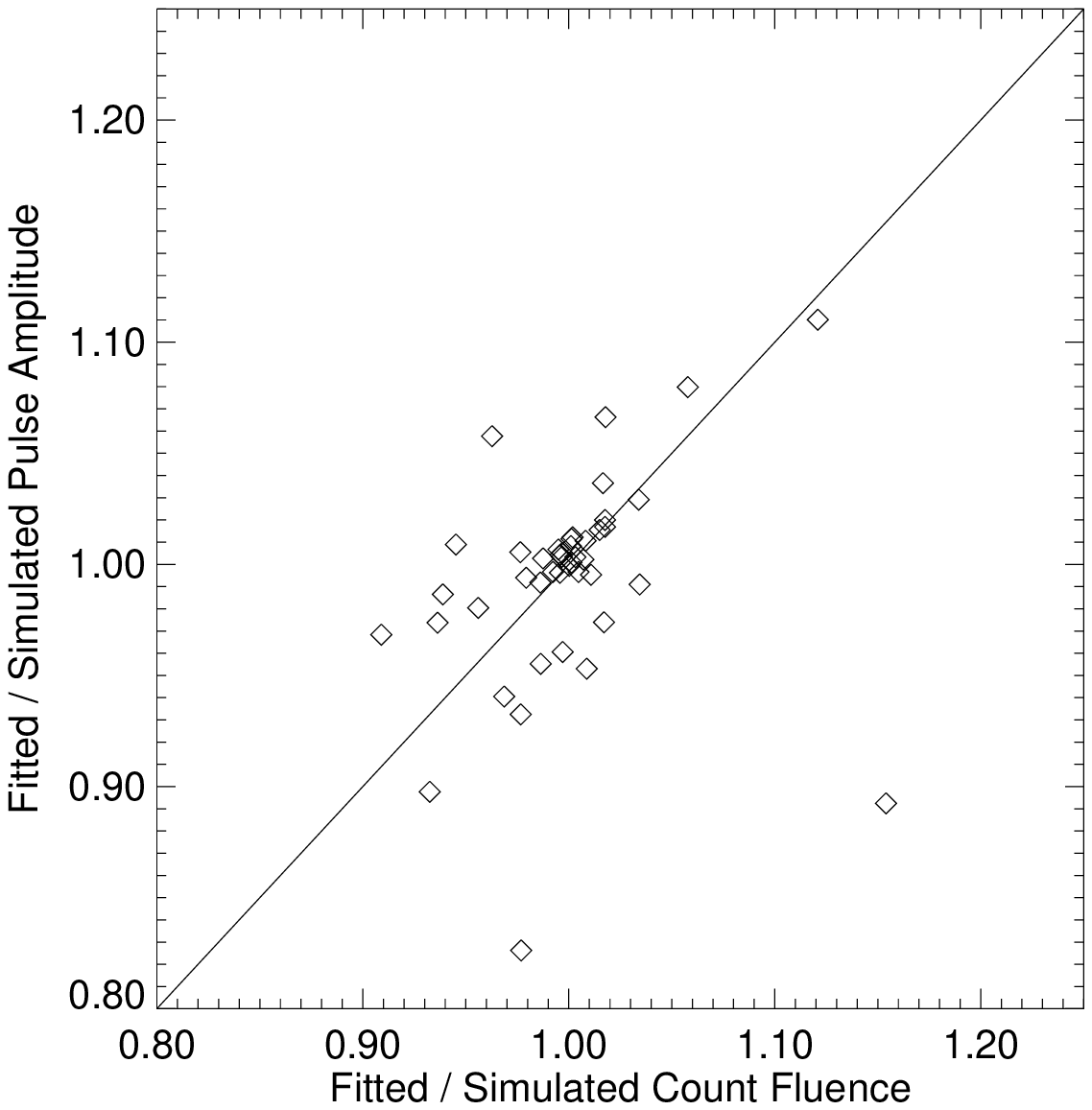}\caption{Ratios of fitted to simulated pulse amplitudes versus ratios of fitted to simulated pulse count fluences, with line of constant pulse width, for single-pulse simulated bursts.  There appears to be a positive correlations between the errors in the
fitted pulse amplitudes and count fluences, but the Spearman
rank-correlation coefficient shows that they are actually
uncorrelated.  \label{sim_rampvsrarea}}
\end{figure}

Figure~\ref{sim_rampvsrarea} shows an apparent positive correlation
between the errors in the fitted pulse amplitudes and fitted count
fluences for simulated bursts consisting of a single pulse in both the
simulation and the fit.  However, the Spearman rank-correlation
coefficient shows no significant correlation between the two sets of
errors.  Therefore, the uncorrelated errors in the pulse amplitudes
and pulse count fluences would tend to smear out any existing
correlations rather than to create correlations, which is what we have
seen above.

\subsection{Correlations Between Pulse Amplitude and Pulse Asymmetry}

It has been reported that when considering the averaged time profiles
of bursts, the decay times from the peaks of bursts show an inverse
correlation with peak flux, while the rise times to the peaks of
bursts show a smaller inverse correlation or no variation at all with
peak flux~\citep{stern:1997,stern:1997b,litvak:1998,stern:1999}.  Such
a result could not come from cosmological time dilation, but would
have to be caused by the burst production mechanism itself, or by some
selection effect, perhaps resulting from the BATSE trigger criteria,
which selects for fast-rising bursts~\citep{higdon:1996}, but is
independent of burst decay times.  It is possible that a similar
effect could appear in the individual pulses comprising a burst, as a
\emph{positive} correlation between pulse amplitudes and pulse
asymmetries as measured by the rise time to decay time ratios.
Although there may be selection effects in the pulse-fitting
procedure, most of these should affect both rise and decay times
similarly, and therefore shouldn't affect pulse asymmetry ratios.

\begin{figure}
\epsfig{file=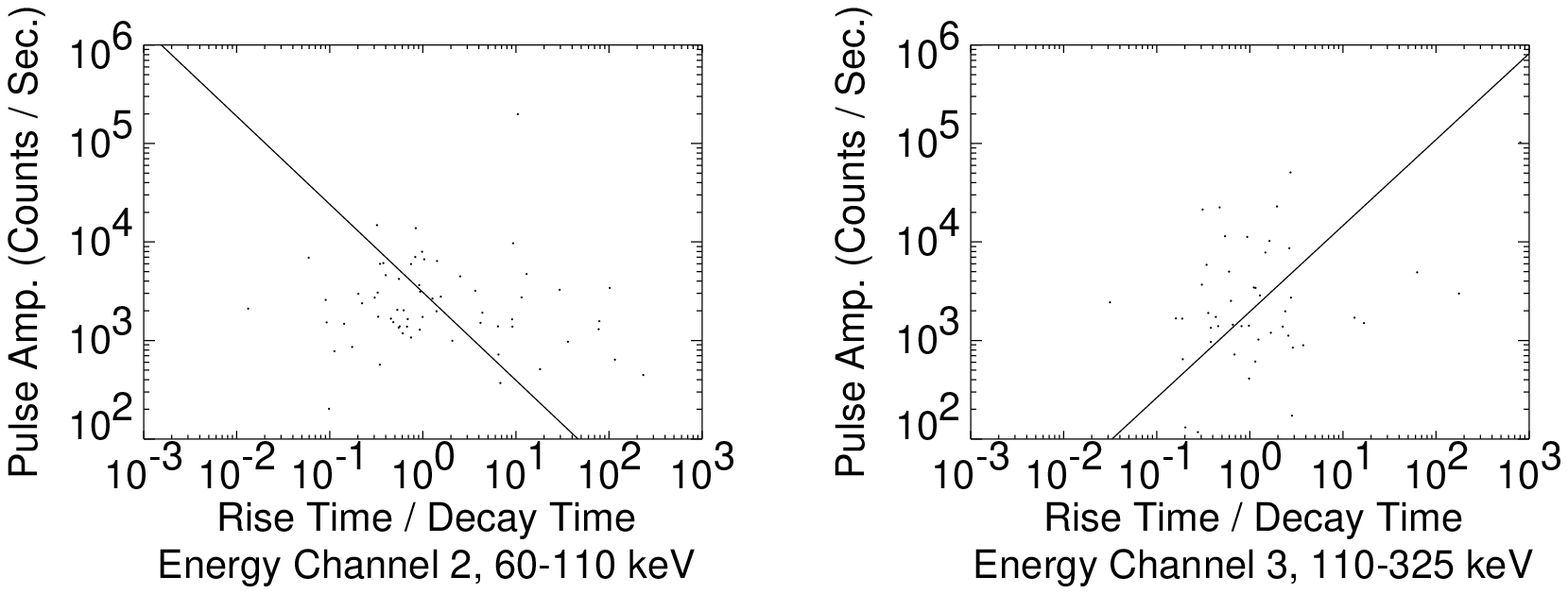}
\epsfig{file=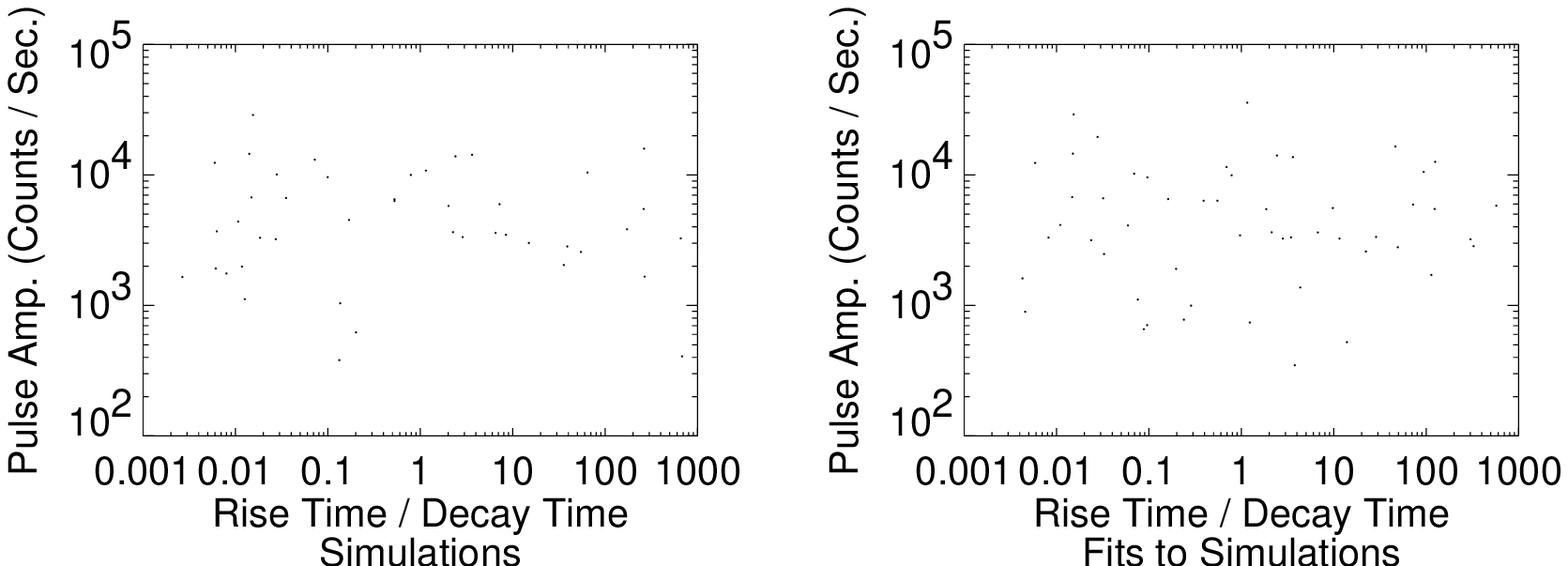}\caption{Pulse amplitude versus pulse asymmetry for single-pulse fits.  In the initial simulations and the fits to the simulations (bottom panels), the correlations were insignificant, so no fits were made. \label{sampvsrd}}
\end{figure}

For bursts consisting of a single pulse, the pulse rise and decay
times are of course the rise and decay times for the entire burst.
Figure~\ref{sampvsrd} shows pulse asymmetries versus pulse amplitudes
for these bursts.  There does not seem to be any clear correlations in
the actual BATSE bursts, but the range of pulse asymmetry ratios
appear to be broader for lower amplitude bursts than for higher
amplitude bursts.  The latter effect could result from the lower
signal-to-noise of lower amplitude pulses.  The Spearman rank-order
correlation coefficients shown in Table~\ref{tab:ampvsrd},
columns~(a), comparing pulse amplitudes and pulse asymmetries of
single-pulse bursts essentially confirm this impression; the
correlations for the actual BATSE bursts are very weak, and have
different signs in the different energy channels.  In the simulated
bursts, there are clearly no correlations in either the initial
simulations or in the fits to the simulations.

\begin{deluxetable}{crlrrlr}
\tablecaption{Correlation Between Pulse Amplitude and Asymmetry for (a) Single Pulse Bursts and (b) Highest Amplitude Pulse in Each Burst, and the Fitted Power Law Index $\delta$. \label{tab:ampvsrd}}
\tablehead{
\colhead{Energy} & \multicolumn{3}{c}{(a) Single Pulse Bursts} & \multicolumn{3}{c}{(b) Highest, All Bursts} \\
\colhead{Channel} & \colhead{$r_{s}$} & \colhead{Prob.} & \colhead{$\delta$} & \colhead{$r_{s}$} & \colhead{Prob.} & \colhead{$\delta$}}
\startdata
1 & -0.25 & 0.049 & $-0.91 \pm 0.10$ & -0.24 & 0.0020 & $-0.89 \pm 0.06$ \\
2 & -0.27 & 0.022 & $-0.89 \pm 0.20$ & -0.32 & \e{1.7}{-5} & $-0.84 \pm 0.06$ \\
3 & 0.01 & 0.93 & $0.87 \pm 0.11$ & -0.15 & 0.051 & $-0.99 \pm 0.04$ \\
4 & -0.27 & 0.12 & $-0.86 \pm 0.20$ & -0.13 & 0.30 & $-0.26 \pm 0.09$ \\
Sim. & 0.11 & 0.42 & \nodata & 0.081 & 0.17 & \nodata \\
Fit to Sim. & -0.0059 & 0.96 & \nodata & 0.066 & 0.27 & \nodata \\
\enddata
\end{deluxetable}

\begin{figure}
\epsfig{file=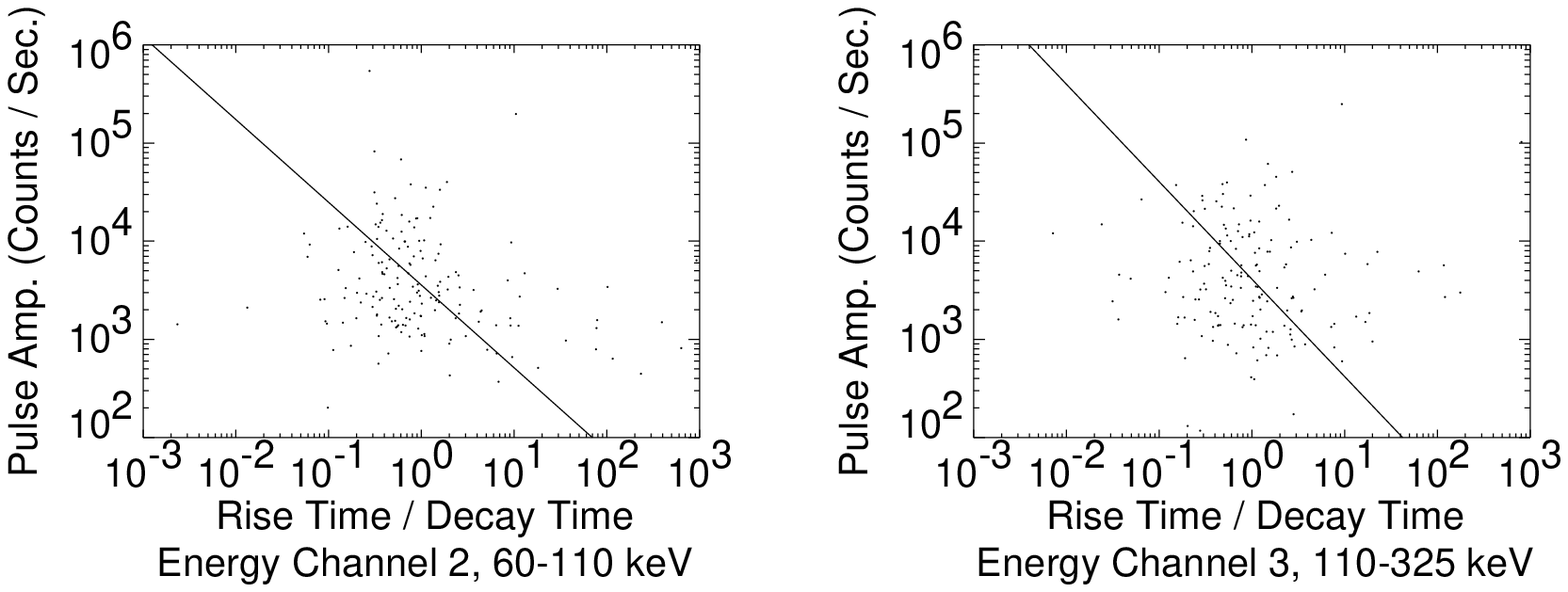}
\epsfig{file=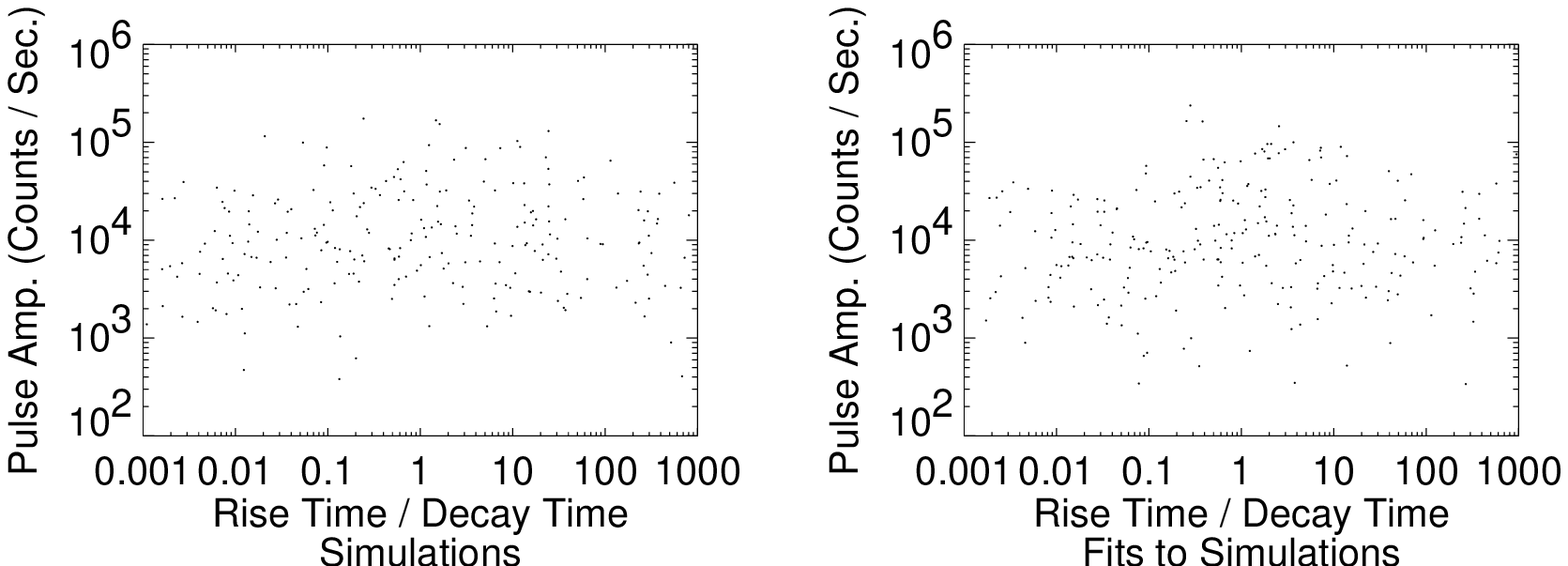}\caption{Pulse amplitude versus pulse asymmetry for highest amplitude pulse in each burst.  In the initial simulations and the fits to the simulations (bottom panels), the correlations were insignificant, so no fits were made. \label{peak_ampvsrd}}
\end{figure}

Although the properties of individual pulses in multiple-pulse bursts
may be different from those of the entire bursts, it may still be
useful to look for correlations between pulse amplitude and asymmetry
for individual pulses in multiple-pulse bursts.  For the highest
amplitude pulses from each burst, plots of pulse asymmetry versus
pulse amplitude are shown in Figure~\ref{peak_ampvsrd}, which again
show that pulse asymmetries span a larger range of values at lower
amplitudes in the actual BATSE bursts.  The Spearman rank-order
correlation coefficients given in Table~\ref{tab:ampvsrd},
columns~(b), show a marginally significant inverse correlations in
energy channels~1 and 3, a strong inverse correlation in channel~2,
and no correlation in channel~4.  Again, the simulated bursts show no
correlation at all.

Finally, we consider correlations between the amplitudes and
asymmetries of pulses within bursts.  Table~\ref{tab:amprd} shows
characteristics of the distributions of the power law indices $\delta$
obtained from fits to these quantities.  There do not appear to be
statistically significant correlations, except possibly in channel 3.

\begin{deluxetable}{crlrrrrr}
\tablecaption{Correlations Between Pulse Asymmetry and Amplitude Within Bursts, and the Distributions and Medians of the Fitted Power Law Index $\delta$. \label{tab:amprd}}
\tablehead{
\colhead{Energy} & \colhead{\% Pos.} & \colhead{Binom.} & \colhead{} & \colhead{} & \colhead{} & \colhead{} & \colhead{Med.} \\
\colhead{Channel} & \colhead{Corr.} & \colhead{Prob.} & \colhead{$\delta < -1$} & \colhead{$-1 < \delta < 0$} & \colhead{$0 < \delta < 1$} & \colhead{$\delta > 1$} & \colhead{$\delta$}}
\startdata
1 & 47/94 = 50\% & 1 & 7 & 42 & 36 & 9 & -0.029 \\
2 & 60/109 = 55\% & 0.29 & 6 & 48 & 44 & 11 & -0.066 \\
3 & 70.5/116 = 61\% & 0.020 & 5 & 52 & 50 & 9 & 0.077 \\
4 & 17/35 = 49\% & 0.87 & 2 & 16 & 14 & 3 & -0.090 \\
Sim. & 109/223 = 49\% & 0.74 & 5 & 106 & 103 & 9 & 0.0068 \\
Fit to Sim. & 100/198 = 51\% & 0.89 & 13 & 81 & 94 & 10 & 0.063 \\
\enddata
\end{deluxetable}

In summary, there is no clear evidence of any correlations between
pulse amplitudes and pulse asymmetry, so that the variations of pulse
rise and decay time with pulse amplitude don't appear to be
significantly different.

\section{Summary and Discussion}
\label{sec:discuss}

In this paper, we use a pulse-fitting procedure to the TTS data from
BATSE and determine the amplitudes, rise and decay times, and
fluences.  We investigate the correlations between all of these
parameters of pulses in individual bursts and among different bursts.
The former gives a measure of correlations intrinsic to the energy and
radiation generation in burst sources, while the latter are also
affected by cosmological effects.  Simulations are used to determine
the biases of the pulse-fitting procedure.

If the peak luminosities of pulses or bursts are approximate standard
candles, so that the peak fluxes would be good measures of distance,
then we expect to find negative correlations between fluxes and
timescales.  We do find inverse correlations between the highest pulse
amplitude within a burst and two different timescales, the width of
the highest amplitude pulse and the time interval between the two
highest amplitude pulses.  The former correlation, between pulse
amplitude and pulse width, which is expected from cosmological time
dilation effects, is nevertheless not consistent with purely
cosmological effects, but must be at least partially influenced by
non-cosmological effects.  These non-cosmological effects may include
intrinsic properties of the burst sources, or selection effects due to
the BATSE triggering procedure, but do not appear to be affected by
the pulse-fitting procedure.  Our study indicates that the latter
correlation, between pulse amplitude and time intervals between
pulses, may be less influenced by non-cosmological effects.  The
inverse correlation observed between pulse amplitude and pulse width
within bursts results in part from selection effects in the
pulse-fitting procedure, but also appears to result in part from
intrinsic properties of the burst sources.

If the total radiated energies of bursts are approximate standard
candles, so that the burst fluences would be good measures of
distance, then we expect to find negative correlations between
fluences and timescales.  We find instead a \emph{positive}
correlation between the total burst count fluence and the width of the
highest amplitude pulse, but no correlation with the time interval
between the two highest amplitude pulses.  The former correlation
indicates that non-cosmological effects are stronger than any
cosmological effects.  This is supported by the positive correlation
between pulse amplitude and pulse count fluence within bursts.
However, it is not clear why total burst count fluence and time
intervals between pulses show no correlation.

It is natural to expect that the peak flux of bursts and the total
count fluence of bursts should both decrease essentially the same way
(except for a factor of $1 + z$) as the distance to the burst sources
increase.  This would suggest that there should be positive
correlations between the peak flux of bursts and the total count
fluence of bursts.  Strangely, the highest pulse amplitude and the
total count fluence of bursts appear to have no statistically
significant correlation with each other, implying that the two
measures of brightness cannot both be good standard candles; at least
one, or more probably both, are poor measures of distance.

There do not appear to be any statistically significant correlations
between pulse amplitude and pulse asymmetry, whether the comparison is
i) of all pulses in all bursts combined, ii) of only the highest pulse
in each burst, iii) of only the single-pulse bursts, or iv) of
different pulses within multiple-pulse bursts.  This implies that the
differences between the variations of pulse rise and decay time with
pulse amplitude are statistically insignificant, and both rise times
and decay times tend to decrease as pulse amplitude increases.

\acknowledgments

We thank Jeffrey Scargle and Jay Norris for many useful discussions.
This work was supported in part by Department of Energy contract
DE--AC03--76SF00515.

{
%\bibliographystyle{abbrvnat}
%\bibstyle@aa
%\bibliographystyle{apj}
%\bibliography{apj-jour,alee}

}

\end{document}